\documentclass[11pt,a4paper]{article}
\usepackage[T1]{fontenc}
\usepackage[utf8]{inputenc}
\usepackage{lmodern}
\usepackage[a4paper,left=23mm,right=23mm,top=22mm,bottom=24mm,footskip=11mm]{geometry}
\usepackage{amsmath,amssymb,bm}
\usepackage{graphicx}
\usepackage{placeins,needspace}
\usepackage{microtype}
\usepackage{titlesec}
\usepackage{etoolbox}
\usepackage[font=small,labelfont=bf,labelsep=period,justification=justified,singlelinecheck=false]{caption}
\usepackage{xurl}
\usepackage[unicode,hidelinks,bookmarksnumbered=true]{hyperref}
\hypersetup{
  pdftitle={Single-photon scattering by a nanofiber-coupled atomic chain: beyond independent radiation},
  pdfauthor={Juliette Jim\'enez-Jaimes and \'Etienne Brion},
  pdfsubject={Author-formatted version of the resubmitted manuscript},
  pdfkeywords={Waveguide quantum electrodynamics, optical nanofiber, single-photon scattering, radiative couplings, temporal modes}
}
\titleformat{\section}{\normalfont\large\bfseries}{\thesection}{0.7em}{}
\titleformat{\subsection}{\normalfont\normalsize\bfseries}{\thesubsection}{0.7em}{}
\titleformat{\paragraph}[runin]{\normalfont\normalsize\bfseries}{}{0pt}{}
\titlespacing*{\section}{0pt}{2.5ex plus 0.5ex minus 0.3ex}{1ex}
\titlespacing*{\subsection}{0pt}{2ex plus 0.4ex minus 0.2ex}{0.6ex}
\titlespacing*{\paragraph}{0pt}{1.5ex plus 0.3ex}{0.7em}
\pretocmd{\paragraph}{\Needspace{6\baselineskip}}{}{}
\newcommand{\ack}[1]{\section*{Acknowledgments}#1}
\newcommand{\funding}[1]{\section*{Funding}#1}
\newcommand{\data}[1]{\section*{Data availability}#1}
\begin{document}
\begin{center}
{\fontsize{18}{22}\selectfont\bfseries Single-photon scattering by a nanofiber-coupled atomic chain: beyond independent radiation\par}
\vspace{4mm}
{\large Juliette Jim\'enez-Jaimes$^1$ and \'Etienne Brion$^{1,*}$\par}
\vspace{2.5mm}
{\small $^1$Laboratoire Collisions Agr\'egats R\'eactivit\'e, UMR 5589, FeRMI,\\
Universit\'e de Toulouse, CNRS, Toulouse, France\par}
\vspace{1.5mm}
{\small $^*$Author to whom any correspondence should be addressed.\\
E-mail: \href{mailto:brion@irsamc.ups-tlse.fr}{brion@irsamc.ups-tlse.fr}\par}
\end{center}
\vspace{2mm}
\begingroup
\small
\noindent\textbf{Abstract}\par\smallskip
\noindent We study single-photon scattering by a nanofiber-coupled atomic chain, comparing a full microscopic model with an independent-radiation approximation. We show that nearly equal transmission probabilities can coexist with substantially different transmitted temporal modes, even after adjusting their relative delay. For a reference chain, a decomposition into collective atomic resonances relates this discrepancy mainly to spectral-phase changes caused by the omitted interatomic couplings. Parameter scans delimit this phase-dominated regime, and controlled comparisons separate the effects of coherent exchange and collective decay.
\par
\endgroup
\vspace{2mm}

\section{Introduction}
Optical nanofibers are produced by heating and stretching a conventional fiber into a subwavelength-diameter waist, connected to the original fiber by gradual, high-transmission tapers~\cite{Ward2014}. The guided field has an evanescent component outside the dielectric that can couple to nearby atoms~\cite{LeKien2005}. It enables two-color trapping, proposed theoretically~\cite{LeKien2004} and demonstrated experimentally~\cite{Vetsch2010}, and controlled atomic transport~\cite{Schneeweiss2013}. Applications include electromagnetically induced transparency, slow light and coherent storage~\cite{Gouraud2015}, and collective Bragg reflection from ordered chains~\cite{Corzo2016}, whose guided response has been studied theoretically~\cite{LeKien2014}. For nearby Rydberg atoms, theory predicts modified emission and level shifts~\cite{Stourm2020}, while excitation through the evanescent field has been demonstrated experimentally~\cite{Vylegzhanin2023}.

Optical nanofibers provide a natural platform for waveguide quantum electrodynamics (WQED), in which guided modes mediate long-range atomic interactions~\cite{AsenjoGreen2017,Kornovan2016} while the array geometry shapes the collective response~\cite{Jimenez2025}.

In the present article, we study single-photon scattering by a nanofiber-coupled atomic chain and compare a reference model~F (for ``full''), which includes guided and radiation-mediated interatomic couplings, with another model~I (for ``independent radiation''), which treats emission away from the fiber as an independent loss at each atom.

Consequences of the omitted couplings are known. Asenjo-Garcia et al.~\cite{AsenjoSelective2017} demonstrated that collective radiation affects photon storage. Liao et al.~\cite{Liao2016} showed that nonguided interactions modify single-photon output fields, while Svendsen and Olmos~\cite{Svendsen2023} calculated nanofiber-modified interactions and their effect on chain transmission, comparing full radiation couplings with their free-space approximation. Here, in contrast, model~I omits the off-diagonal radiation couplings altogether. Collective transients have also been studied~\cite{Pennetta2022,Cardenas2023,Su2023}. Comparing transfer-matrix and microscopic calculations for stratified atomic media, Sokolov and Guerin showed how interlayer coupling at high density invalidates a description based on separate local responses~\cite{Sokolov2026}.

Our goal here is to quantify how much the transmitted photon modes can differ when the two models predict nearly equal transmission probabilities. The modes are defined by their normalized complex spectral amplitudes. Equal integrated transmission does not, in general, determine the output of a linear filter. Our contribution is to quantify the discrepancy produced by a specific microscopic approximation after removal of a global phase and an optimal relative delay, and to connect it to identified collective resonances. We also determine the parameter region where close transmission predictions coexist with appreciable mode mismatch and separate the effects of coherent and dissipative radiation couplings.

\section{Physical system and scattering descriptions}
In this section, we specify the geometry and atom--field equations, define the two interaction models and derive their frequency-dependent scattering amplitudes.

\begin{figure}[!htb]
\centering\includegraphics[width=\textwidth]{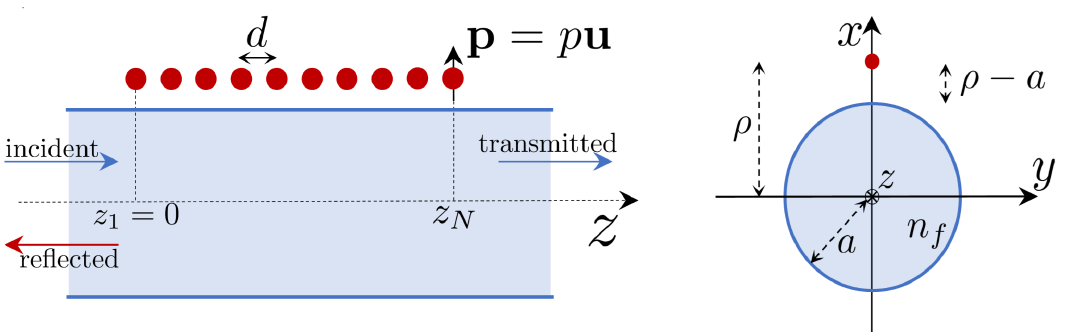}
\caption{Chain of $N=10$ atoms near an optical nanofiber. Left: longitudinal view showing the spacing $d$, end positions $z_1=0$ and $z_N$, and incident, transmitted and reflected guided fields. The dipole $\mathbf p=p\mathbf u$ is radial, as in the numerical example. Right: transverse view showing the fiber radius $a$, refractive index $n_f$, atomic distance $\rho$ from the axis and atom--surface distance $\rho-a$. Blue shading represents the dielectric.}
\label{fig:system}
\end{figure}

\begin{samepage}
\subsection{Geometry and assumptions}
We model the nanofiber as an infinite, lossless dielectric cylinder of radius $a$ and refractive index $n_f$ in vacuum, excluding the distant tapers. Its axis is chosen as the Cartesian $z$ axis. The $N$ atoms form a parallel chain (Fig.~\ref{fig:system}), with Cartesian positions
\begin{equation}
 \mathbf r_j=(\rho,0,z_j),\qquad z_j=(j-1)d,\qquad j=1,\ldots,N.
 \label{eq:geometry}
\end{equation}
\end{samepage}
Here $\rho$ is the distance from the axis, $d$ the longitudinal separation and $\rho-a$ the atom--surface distance. The atoms have identical two-level transitions of angular frequency $\omega_a$. Their real transition dipole is $p\mathbf u$, with magnitude $p$ and common radial unit vector $\mathbf u$.

For an incident field component of angular frequency $\omega$, we define the detuning $\delta=\omega-\omega_a$. We also define $k_a=\omega_a/c$ and $\lambda_a=2\pi/k_a$, with $c$ the speed of light in vacuum.

Throughout this article, we shall assume the single-mode regime in which the fiber supports only the fundamental HE$_{11}$ spatial mode with two orthogonal polarizations and two propagation directions~\cite{LeKien2005,LeKien2014}. At the common atomic azimuth, the dipoles couple to one quasilinear polarization, the other being decoupled. The incident photon travels along $+z$. The propagation constant $\beta\equiv\beta(\omega_a)>0$ is the HE$_{11}$ mode's longitudinal wave number at resonance. With time dependence $e^{-{\mathrm{i}}\omega t}$, forward and backward spatial phases are $e^{{\mathrm{i}}\beta z}$ and $e^{-{\mathrm{i}}\beta z}$. The atoms are assumed fixed and initially prepared in their ground states.

\subsection{Atom--field equations}
\label{sec:couplings}
The dynamics of the coupled atom--field system may be studied following the same steps as in free space. First, one derives coupled Heisenberg equations for the field and atomic operators from the full Hamiltonian, including the free Hamiltonians of the atoms and the medium-assisted field, together with their electric-dipole interaction~\cite{ScheelBuhmann2008}. Formally integrating the field equations and substituting their solutions into the atomic equations then yields a self-consistent set of integro-differential equations for the atomic operators.

We assume weak coupling and near-resonant excitation, with the relevant detunings, collective frequency shifts and decay rates much smaller than $\omega_a$. After factoring out the rapid optical oscillation at $\omega_a$, the atomic operators therefore have slowly varying envelopes. The Markov approximation further requires the reservoir correlation time and the light-propagation times between atoms to be short compared with the evolution timescales of these envelopes and with the incident pulse duration. The delayed atomic envelopes in the memory integrals can then be replaced by their values at the current time, yielding equations that are local in time.

We also assume that the electromagnetic response of the nanofiber environment in the absence of atoms varies negligibly over the frequency range covering the incident pulse and the relevant collective resonances. The electromagnetic factors entering the couplings are therefore evaluated at $\omega_a$. This neglects propagation delays on the envelope timescale, but preserves the spatial phases accumulated at the optical frequency, including the guided propagation factors $e^{\mathrm{i}\beta(\omega_a)|z_i-z_j|}$. The corresponding phases need not be small: the atomic separations are not assumed negligible compared with the wavelength. Evaluating the electromagnetic response at a fixed frequency does not suppress the frequency dependence of the atomic response, which still exhibits its collective resonances and their associated dynamics.

After these steps have been completed, one keeps only the resonant terms in the effective dynamics. In a frame rotating at $\omega_a$, also used for the field envelopes below, one gets the following dynamical equation for the atomic operators~\cite{AsenjoGreen2017,Jimenez2025}
\begin{equation}
 \dot{\hat\sigma}_i^-
 =-\mathrm i\sum_j\left(\Omega_{ij}-\frac{\mathrm i}{2}\Gamma_{ij}\right)
 \hat\sigma_i^z\hat\sigma_j^-
 +\frac{\mathrm ip}{\hbar}\hat\sigma_i^z\,
 \mathbf u\cdot\hat{\mathbf E}^{(+)}_{\rm in}(\mathbf r_i,t),
 \label{eq:atomic-operators}
\end{equation}
where $\hat\sigma_i^-=|g\rangle_i\langle e|$, $\hat\sigma_i^+=(\hat\sigma_i^-)^\dagger$ and $\hat\sigma_i^z=|g\rangle_i\langle g|-|e\rangle_i\langle e|$. Moreover, the field satisfies the input--output relation
\begin{equation}
 \hat{\mathbf E}^{(+)}(\mathbf r,t)
 =\hat{\mathbf E}^{(+)}_{\rm in}(\mathbf r,t)
 +\mu_0\omega_a^2p\sum_j\mathbf G(\mathbf r,\mathbf r_j,\omega_a)
 \cdot\mathbf u\,\hat\sigma_j^-(t).
 \label{eq:field-input-output}
\end{equation}
The input evolves in the nanofiber environment as if the atoms were absent and includes vacuum fluctuations. The source term sums the fields at the observation point $\mathbf r$ induced by the atomic dipoles $p\mathbf u\hat\sigma_j^-$ at $\mathbf r_j$. Here $\mathbf G$ is the outgoing Green tensor of Maxwell's equations in the presence of the fiber, including free-space and dielectric scattering contributions~\cite{ScheelBuhmann2008}. The same tensor determines the atomic couplings in Eq.~\eqref{eq:atomic-operators}:
\begin{equation}
 \begin{aligned}
 \Omega_{ij}&=-\frac{\mu_0\omega_a^2p^2}{\hbar}\,
 \mathbf u\cdot\operatorname{Re}\mathbf G(\mathbf r_i,\mathbf r_j,\omega_a)\cdot\mathbf u,
 \qquad i\ne j,\\
 \Gamma_{ij}&=\frac{2\mu_0\omega_a^2p^2}{\hbar}\,
 \mathbf u\cdot\operatorname{Im}\mathbf G(\mathbf r_i,\mathbf r_j,\omega_a)\cdot\mathbf u.
 \end{aligned}
 \label{eq:green}
\end{equation}
The off-diagonal elements of the matrices $\bm\Omega=[\Omega_{ij}]$ and $\bm\Gamma=[\Gamma_{ij}]$ describe coherent excitation exchange and collective emission, respectively. Both matrices are real symmetric for this reciprocal geometry and real dipoles. The common single-atom shift, including the fiber-induced contribution, is absorbed into $\omega_a$, so $\Omega_{jj}=0$; $\Gamma_{jj}=\gamma$ is the single-atom population-decay rate. Thus $\omega_a$ refers to the transition of an isolated atom at the chosen distance from the fiber. Note that the coherent coupling includes second-order virtual processes of the full dipole interaction, including counter-rotating contributions omitted by the rotating-wave approximation~\cite{Dung2002,Kornovan2016,Miroshnychenko2013,Derouault2014}.

These couplings also define the non-Hermitian effective Hamiltonian~\cite{AsenjoGreen2017,Jimenez2025}, which, for vacuum input, generates the nonunitary evolution between quantum jumps in Monte Carlo wave-function simulations of the master equation for the reduced atomic density operator~\cite{Dalibard1992}:
\begin{equation}
 \hat H_{\rm eff}=\hbar\sum_{i,j}\left(\Omega_{ij}-\frac{\mathrm i}{2}\Gamma_{ij}\right)
 \hat\sigma_i^+\hat\sigma_j^-.
 \label{eq:generator}
\end{equation}
In the single-excitation basis $|e_j\rangle=\hat\sigma_j^+|G\rangle$, with $|G\rangle$ the collective atomic ground state, the matrix of $\hat H_{\rm eff}$ is denoted by $\bm H_{\rm eff}$, so $\bm H_{\rm eff}/\hbar=\bm\Omega-\mathrm i\bm\Gamma/2$. Its right eigenstates in this subspace are called radiative eigenmodes (REMs); their complex eigenvalues determine the collective frequency shifts and decay rates analyzed in Sec.~\ref{sec:collective}.

\subsection{Guided and radiation-mediated couplings}
\label{sec:models}
The decomposition $\mathbf G=\mathbf G_g+\mathbf G_{\mathrm{rad}}$ separates guided and nonguided contributions. We underline that, here, $\mathbf G_{\mathrm{rad}}$ includes both vacuum propagation and its modification by the fiber. The single-atom population-decay rate hence splits as $\gamma=\gamma_g+\gamma_{\mathrm{rad}}$.

The transverse dipoles emit equally into the two coupled guided directions, at rate $\gamma_g/2$ each. The guided contribution to the effective Hamiltonian is~\cite{LeKien2014,AsenjoGreen2017}
\begin{equation}
 \frac{(H_{{\rm eff},g})_{ij}}{\hbar}=-\frac{\mathrm i\gamma_g}{2}e^{\mathrm i\beta|z_i-z_j|},
 \label{eq:guided}
\end{equation}
with $(\Omega_g)_{ij}=(\gamma_g/2)\sin(\beta|z_i-z_j|)$ and $(\Gamma_g)_{ij}=\gamma_g\cos(\beta|z_i-z_j|)$. The radiation contribution is $\bm H_{{\rm eff},{\mathrm{rad}}}/\hbar=\bm\Omega_{\mathrm{rad}}-\mathrm i\bm\Gamma_{\mathrm{rad}}/2$. Its off-diagonal entries and decay rates follow from $\mathbf G_{\mathrm{rad}}$ in Eq.~\eqref{eq:green}~\cite{Kornovan2016,Svendsen2023,Jimenez2025}. As in Sec.~\ref{sec:couplings}, the common single-atom frequency shift, including the fiber-induced shift, is absorbed into $\omega_a$; hence $(\Omega_{\rm rad})_{jj}=0$, while $(\Gamma_{\rm rad})_{jj}=\gamma_{\rm rad}$. Both decay matrices are positive semidefinite: for any normalized state in the single-excitation subspace, their expectation values give the nonnegative emission rates into the guided and radiation modes, respectively~\cite{ScheelBuhmann2008,AsenjoGreen2017}.

We compare the full model~F with the independent-radiation model~I, respectively described by the Hamiltonians
\begin{equation}
 \begin{aligned}
 \bm H_{{\rm eff},F}&=\bm H_{{\rm eff},g}+\bm H_{{\rm eff},{\mathrm{rad}}},\\
 \bm H_{{\rm eff},I}&=\bm H_{{\rm eff},g}-\frac{\mathrm i\hbar\gamma_{\mathrm{rad}}}{2}\mathbb I,
 \end{aligned}
 \label{eq:models}
\end{equation}
where $\mathbb I$ is the $N\times N$ identity. Model~F includes the full guided and radiation contributions while model~I removes the off-diagonal radiation couplings. Both models share identical local atomic properties and account for all guided multiple scattering. Thus $\bm\Omega_I=\bm\Omega_g$ and $\bm\Gamma_I=\bm\Gamma_g+\gamma_{\mathrm{rad}}\mathbb I$. For the atomic chain considered, model~I is fully equivalent to the single-atom transfer-matrix approach~\cite{LeKien2014}.

\subsection{Single-photon scattering amplitudes}
\label{sec:scattering}
This section sketches the main steps of the calculation of single-photon scattering amplitudes and the associated flux balance. Step-by-step derivations may be found in Appendix~\ref{app:scattering}.

For the incident state $|\Psi_{\rm in}\rangle=|G;1_{\rm in}\rangle$, with all atoms in the ground state, one incoming guided photon and all other photonic modes in vacuum, we introduce the atomic excitation amplitudes
\begin{equation*}
 \alpha_j(t)=\langle G;0|\hat\sigma_j^-(t)|\Psi_{\rm in}\rangle,
\end{equation*}
which give the probability amplitudes for atom $j$ to be excited at time $t$, with all other atoms in their ground states and the field in vacuum.

Let $f(\delta)$ be the incident single-photon spectral amplitude, normalized by $\int|f(\delta)|^2{\mathrm d}\delta=1$. Its temporal amplitude and the Fourier transforms of the atomic amplitudes are defined by
\begin{equation*}
 \xi_{\rm in}(t)=\frac{1}{\sqrt{2\pi}}\int f(\delta)e^{-\mathrm i\delta t}{\mathrm d}\delta,
 \qquad
 \bar{\bm\alpha}_X(\delta)=\frac{1}{\sqrt{2\pi}}\int\bm\alpha_X(t)e^{\mathrm i\delta t}{\mathrm d}t.
\end{equation*}
Detuning integrals run over $\mathbb R$, the narrowband extension of the physical range $\delta>-\omega_a$. The incident guided field is normalized by
\begin{equation*}
 -\frac{p}{\hbar}\mathbf u\cdot
 \langle G;0|\hat{\mathbf E}^{(+)}_{\rm in}(\mathbf r_j,t)|\Psi_{\rm in}\rangle
 =\sqrt{\frac{\gamma_g}{2}}\,\xi_{\rm in}(t)e^{\mathrm i\beta z_j},
\end{equation*}
so that $|\xi_{\rm in}(t)|^2$ is the incident photon flux, referred to $z=0$. The atom--mode coupling at $z=0$ is chosen real and positive. Taking the same matrix elements of Eq.~\eqref{eq:atomic-operators} and Fourier transforming in time gives the atomic spectral response~\cite{Liao2016}
\begin{equation}
 \begin{aligned}
 \bar{\bm\alpha}_X(\delta)&=\sqrt{\frac{\gamma_g}{2}}\,f(\delta)
 \bm M_X^{-1}(\delta)\mathbf b,\\
 \bm M_X(\delta)&=\delta\mathbb I-\frac{\bm H_{{\rm eff},X}}{\hbar},
 \qquad b_j=e^{\mathrm i\beta z_j},
 \end{aligned}
 \label{eq:solve}
\end{equation}
for $X=F,I$, where $\mathbf b$ contains the incident phases.

Projecting Eq.~\eqref{eq:field-input-output} onto the coupled guided modes selects only $\mathbf G_g$ in the source term. The outgoing amplitudes are evaluated outside the chain and referred to $z=0$ by removing empty-fiber propagation times and phases. Writing the transmitted and reflected spectra as $f(\delta)t_X(\delta)$ and $f(\delta)r_X(\delta)$ gives~\cite{LeKien2014,AsenjoGreen2017}
\begin{equation}
 \begin{aligned}
 t_X(\delta)&=1-\frac{\mathrm i\gamma_g}{2}\mathbf b^\dagger\bm M_X^{-1}(\delta)\mathbf b,\\
 r_X(\delta)&=-\frac{\mathrm i\gamma_g}{2}\mathbf b^{\mathsf T}\bm M_X^{-1}(\delta)\mathbf b.
 \end{aligned}
 \label{eq:rt}
\end{equation}
Superscripts $\dagger$ and $\mathsf T$ denote Hermitian conjugation and transpose without conjugation. The factors $e^{\mp\mathrm i\beta z_j}$ describe emission into the forward and backward directions; transmission also contains the directly propagated incident field. Each input or output coupling contributes a factor $\sqrt{\gamma_g/2}$.

With this normalization, $|f(\delta)|^2|t_X(\delta)|^2$ and $|f(\delta)|^2|r_X(\delta)|^2$ are the transmitted and reflected spectral probability densities. The corresponding radiation density is $|f(\delta)|^2L_X(\delta)$, where (Appendix~\ref{app:radiation})
\begin{equation*}
 L_X(\delta)=\frac{\gamma_g}{2}\mathbf b^\dagger
 [\bm M_X^{-1}(\delta)]^\dagger\bm\Gamma_{{\mathrm{rad}},X}
 \bm M_X^{-1}(\delta)\mathbf b,
\end{equation*}
with $\bm\Gamma_{{\mathrm{rad}},F}=\bm\Gamma_{\mathrm{rad}}$ and $\bm\Gamma_{{\mathrm{rad}},I}=\gamma_{\mathrm{rad}}\mathbb I$. Positivity ensures $L_X(\delta)\geq0$. As shown in Appendix~\ref{app:flux}, the transmission, reflection and radiation-loss coefficients satisfy probability conservation at each detuning:
\begin{equation}
 |t_X(\delta)|^2+|r_X(\delta)|^2+L_X(\delta)=1.
 \label{eq:flux}
\end{equation}

\section{Transmission probability and transmitted mode}
\label{sec:pulse}
Integrating the transmitted spectral probability density gives
\begin{equation}
 P_{T,X}=\int |f(\delta)|^2|t_X(\delta)|^2{\mathrm{d}}\delta.
 \label{eq:probability}
\end{equation}
Replacing $|t_X|^2$ by $|r_X|^2$ or $L_X$ gives the reflection probability $P_{R,X}$ or radiative-loss probability $P_{L,X}$, respectively, with $P_{T,X}+P_{R,X}+P_{L,X}=1$.

For $P_{T,X}>0$, conditioning on transmission gives the normalized spectral amplitude and temporal envelope
\begin{equation}
 \begin{aligned}
 \psi_X(\delta)&=\frac{f(\delta)t_X(\delta)}{\sqrt{P_{T,X}}},\qquad X=F,I,\\[3pt]
 \widetilde\psi_X(t)&=\frac{1}{\sqrt{2\pi}}
 \int\psi_X(\delta)e^{-{\mathrm{i}}\delta t}{\mathrm{d}}\delta.
 \end{aligned}
 \label{eq:mode}
\end{equation}
Normalization separates the mode from its probability of occurrence. With the propagation convention above, $|\widetilde\psi_X(t)|^2$ is the conditional arrival-time density at an ideal detector beyond the chain. Transmission leaves the atoms in their ground state and all other field channels in vacuum, so the conditional photon state is pure under the stated assumptions. A discrepancy between the transmitted modes therefore demonstrates a mode-prediction error relative to model~F.

The overlap of two normalized temporal amplitudes provides a natural measure of their similarity, taking both magnitude and phase into account. Its squared modulus is insensitive to an overall phase. To distinguish pulse-shape changes from a simple difference in arrival time, we allow the pulse predicted by model~I to be delayed by $\tau$ and define
\begin{equation}
 \begin{aligned}
 {\mathcal{F}}(\tau)&=\left|\int\widetilde\psi_F^*(t)
 \widetilde\psi_I(t-\tau){\mathrm{d}} t\right|^2\\
 &=\left|\int\psi_F^*(\delta)\psi_I(\delta)
 e^{{\mathrm{i}}\delta\tau}{\mathrm{d}}\delta\right|^2.
 \end{aligned}
 \label{eq:fidelity}
\end{equation}
Positive $\tau$ delays the envelope of model~I. The comparison at a common time origin and the best agreement after temporal alignment are quantified by
\begin{equation}
 {\mathcal{F}}_0\equiv{\mathcal{F}}(0),\qquad
 {\mathcal{F}}_{\rm d}\equiv\max_{\tau\in\mathbb R}{\mathcal{F}}(\tau).
 \label{eq:fidelity-measures}
\end{equation}
The optimum delay $\tau_*$ gives ${\mathcal{F}}_{\rm d}={\mathcal{F}}(\tau_*)$, with ${\mathcal{F}}_0\leq{\mathcal{F}}_{\rm d}\leq1$. It equals unity exactly when the two modes coincide up to delay and global phase.

To distinguish spectral-modulus changes from residual spectral-phase differences, we also use
\begin{equation}
 {\mathcal{F}}_{\rm abs}=\left(\int|\psi_F(\delta)||\psi_I(\delta)|{\mathrm{d}}\delta\right)^2.
 \label{eq:abs}
\end{equation}
\begin{samepage}
For any delay, the triangle inequality bounds the modulus in Eq.~\eqref{eq:fidelity} by $\int|\psi_F||\psi_I|\,\mathrm d\delta$, giving $\mathcal F_{\rm d}\leq\mathcal F_{\rm abs}$. The Cauchy--Schwarz inequality then gives
\begin{equation*}
 \mathcal F_{\rm abs}
 \leq\left(\int|\psi_F(\delta)|^2\,\mathrm d\delta\right)
       \left(\int|\psi_I(\delta)|^2\,\mathrm d\delta\right)=1,
\end{equation*}
because each conditional spectral amplitude is normalized to unit probability. The upper bound ${\mathcal{F}}_{\rm abs}$ is attained if an arbitrary frequency-dependent relative phase can be removed. Its gap from ${\mathcal{F}}_{\rm d}$ therefore measures the additional mismatch that no single delay compensates.
\end{samepage}

\subsection{Operational meaning of the mode overlap}
\label{sec:operational}
After optimal temporal alignment, the normalized state predicted by model~F can be decomposed into the mode predicted by model~I and a state orthogonal to it. Choosing the overall phase of the latter mode appropriately gives
\begin{equation}
 |1_{\psi_F}\rangle=\sqrt{\mathcal F_{\rm d}}\,|1_{\psi_I,\tau_*}\rangle
 +\sqrt{1-\mathcal F_{\rm d}}\,|1_\perp\rangle .
 \label{eq:operational}
\end{equation}
Equation~\eqref{eq:operational} can be interpreted through a single-photon mode-selection thought experiment. Consider a hypothetical device placed after the atomic chain that directs the mode predicted by model~I, after optimal temporal alignment, to an accepted output, and all orthogonal modes to other outputs. Only the transmitted photon enters this device. For a photon in the mode predicted by model~F, the acceptance and rejection probabilities are respectively $\mathcal F_{\rm d}$ and $1-\mathcal F_{\rm d}$, conditional on transmission through the chain. The probability of reaching the accepted output per photon incident on the chain is therefore $P_{T,F}\mathcal F_{\rm d}$.

A second, distinct thought experiment probes the same overlap through two-photon interference. Consider a transmitted photon in the pure mode $\psi_F$ entering one input of an ideal balanced beam splitter, and an independently prepared pure reference photon in the mode $\psi_I$ entering the other. The photons have the same polarization and spatial mode at the beam splitter, and their relative arrival time is adjustable. A detector at each output records coincidences, meaning one photon detected in each output during the same trial. Conditional on both photons reaching the beam splitter, and integrating over their full temporal profiles, the coincidence probability at delay $\tau$ is $[1-\mathcal F(\tau)]/2$~\cite{Legero2003}. Its minimum is therefore $(1-\mathcal F_{\rm d})/2$. Perfectly matching modes give no coincidences: the photons leave together through the same output. Orthogonal temporal modes instead give a coincidence probability of $1/2$.

The bound $\mathcal F_{\rm abs}$ answers a different question: how well could the modes match if their relative phase were corrected independently at every frequency, without changing either spectral modulus? A delay only corrects a phase linear in frequency, whereas this hypothetical correction removes all relative spectral-phase variation. Thus, $\mathcal F_{\rm abs}$ is the maximum squared overlap obtainable by changing only the relative spectral phase, while leaving both spectral moduli unchanged.

\section{A microscopic numerical example}
\subsection{Parameters and calculation}\label{sec:parameters}
Our reference chain has radial dipoles and
\begin{equation}
 N=10,\quad k_a a=1.4,\quad n_f=1.45,\quad
 \rho=2a,\quad d=0.3\lambda_a.
 \label{eq:parameters}
\end{equation}
The single-atom rates are $\gamma_g=0.1279\gamma_0$ and $\gamma_{\mathrm{rad}}=0.9295\gamma_0$, where $\gamma_0$ is the spontaneous-emission rate of an isolated atom in vacuum. The HE$_{11}$ propagation constant is $\beta=1.054k_a$.

We use a Gaussian input whose spectral intensity has mean detuning $\delta_c$ and standard deviation $\sigma$,
\begin{equation}
 |f(\delta)|^2=\frac{e^{-(\delta-\delta_c)^2/(2\sigma^2)}}{\sqrt{2\pi}\sigma},
 \qquad \delta_c=-\gamma_0,\quad \sigma=0.5\gamma_0.
 \label{eq:gaussian}
\end{equation}
We choose $f(\delta)$ real and positive, giving the temporal amplitude
\begin{equation*}
 \xi_{\rm in}(t)=\left(\frac{2\sigma^2}{\pi}\right)^{1/4}
 e^{-\sigma^2t^2-{\mathrm{i}}\delta_ct}.
\end{equation*}
Its intensity is centered at $t=0$ with root-mean-square duration $1/(2\sigma)$. Its carrier $\omega_a+\delta_c$ lies below resonance.

We evaluate the response~\eqref{eq:solve} across the input spectrum and optimize the relative delay. Appendix~\ref{app:methods} specifies the cylindrical Green-tensor calculation, tabulates all distinct radiation couplings, and reports convergence tests at both the electromagnetic and pulse-integration levels. For the reference case, these tests support conservative numerical accuracy estimates of $10^{-7}$ in the probabilities and overlaps and $10^{-6}/\gamma_0$ in the optimum delay. These estimates concern numerical discretization, not uncertainty in the physical model or atomic positions.

\subsection{Nearly equal transmission probabilities, different transmitted modes}\label{sec:reference-result}
The transmission probabilities in Table~\ref{tab:probabilities} differ by only $0.34\%$ relative to $P_{T,F}$; reflection and radiation loss show larger differences. Nevertheless, ${\mathcal{F}}_0=0.5848$ and ${\mathcal{F}}_{\rm d}=0.6218$ at $\tau_*=0.9136/\gamma_0$: a substantial mode discrepancy remains after temporal alignment. In the operational terms of Sec.~\ref{sec:operational}, about $38\%$ of the transmitted probability lies outside the mode predicted by~I. An ideal projection onto that mode would succeed with probability $P_{T,F}\mathcal F_{\rm d}=0.3437$ per incident photon, instead of $P_{T,I}=0.5508$ predicted by~I.

\begin{table}[!htb]
\centering
\caption{Probabilities per incident photon for the Gaussian pulse in Eq.~\eqref{eq:gaussian}. All three channels are integrated over the same input spectrum; their probabilities sum to unity before rounding.}
\label{tab:probabilities}
\begin{tabular}{lrr}
\hline
Output channel & Model~F & Model~I\\
\hline
Transmission & 0.5527 & 0.5508\\
Reflection & 0.01816 & 0.002926\\
Radiation away from the fiber & 0.4291 & 0.4462\\
\hline
\end{tabular}
\end{table}

\begin{figure}[!htb]
\centering\includegraphics[width=.495\textwidth]{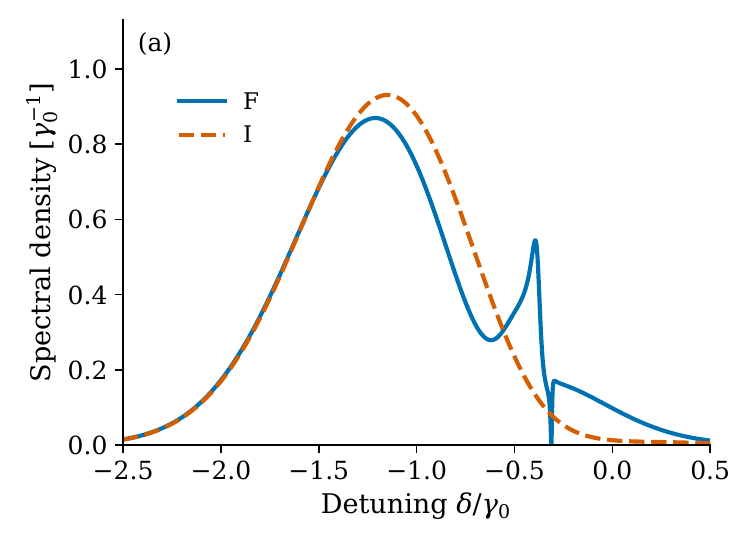}\hfill\includegraphics[width=.495\textwidth]{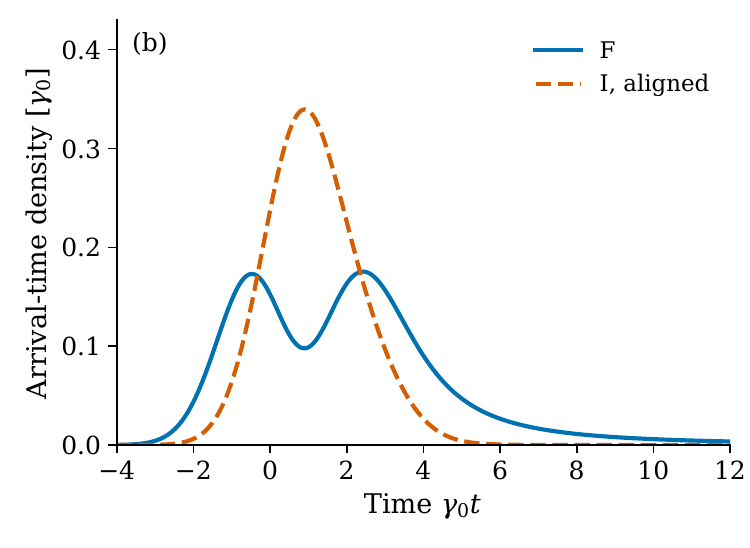}
\caption{Transmitted photon for the chain and input pulse in Eqs.~\eqref{eq:parameters} and \eqref{eq:gaussian}. (a) Normalized spectral densities $|\psi_F|^2$ and $|\psi_I|^2$, in units of $\gamma_0^{-1}$. (b) Conditional arrival-time densities $|\widetilde\psi_F(t)|^2$ and $|\widetilde\psi_I(t-\tau_*)|^2$, in units of $\gamma_0$, with $\tau_*=0.9136/\gamma_0$. Blue solid curves represent~F and orange dashed curves represent~I; line styles also distinguish the curves in grayscale. The origin $t=0$ is the peak of the incident Gaussian, so the shifted~I trace has its own peak displaced from this origin. Each curve is normalized over the entire spectrum or all time, not only over the displayed interval. The two maxima in~(b) belong to one conditional one-photon wave packet, not to two separate emitted photons.}
\label{fig:modes}
\end{figure}

The spectral-modulus overlap, ${\mathcal{F}}_{\rm abs}=0.9641$, is much larger than ${\mathcal{F}}_{\rm d}$. The gap $\mathcal F_{\rm abs}-\mathcal F_{\rm d}=0.3422$ accounts for most of $1-\mathcal F_{\rm d}=0.3782$, whereas $1-\mathcal F_{\rm abs}=0.03591$. Relative spectral phase therefore dominates the mismatch for this pulse. Note that this comparison of overlaps is not a claim that phase and modulus arise from independent microscopic mechanisms. Figure~\ref{fig:modes} shows that the conditional arrival-time distributions also differ after optimal temporal alignment.

\FloatBarrier

The two-lobed profile in Fig.~\ref{fig:modes}(b) results from interference between the directly transmitted pulse and the field re-emitted by the atoms. To calculate these contributions, we use the Gaussian amplitude $\xi_{\rm in}(t)$ given in Sec.~4.1 and the transmission coefficient $t_F(\delta)$ of Eq.~\eqref{eq:rt}. Before conditioning on transmission, the outgoing temporal amplitude and its scattered part are
\begin{equation*}
 \begin{aligned}
 \xi_{T,F}(t)&=\frac{1}{\sqrt{2\pi}}\int f(\delta)t_F(\delta)e^{-\mathrm i\delta t}\,\mathrm d\delta,\\
 \xi_{\rm sc}(t)&=\frac{1}{\sqrt{2\pi}}\int f(\delta)[t_F(\delta)-1]e^{-\mathrm i\delta t}\,\mathrm d\delta,
 \end{aligned}
\end{equation*}
so that $\xi_{T,F}=\xi_{\rm in}+\xi_{\rm sc}$. All three amplitudes use the same reference plane and time origin. Their units are the square root of a photon flux, and $\int|\xi_{T,F}(t)|^2\,\mathrm dt=P_{T,F}$. Thus $|\xi_{T,F}(t)|^2\,\mathrm dt$ is the probability per incident photon of transmission and detection in $[t,t+\mathrm dt]$. Conditional on transmission, the arrival-time probability density plotted in Fig.~\ref{fig:modes}(b) is $|\widetilde\psi_F(t)|^2=|\xi_{T,F}(t)|^2/P_{T,F}$; its integral over time is unity. Expanding the squared modulus gives
\begin{equation}
 P_{T,F}|\widetilde\psi_F(t)|^2
 =|\xi_{\rm in}(t)|^2+|\xi_{\rm sc}(t)|^2
 +2\operatorname{Re}[\xi_{\rm in}^*(t)\xi_{\rm sc}(t)].
 \label{eq:two-lobes}
\end{equation}
The last term is the interference contribution. The first two terms are not probabilities for mutually exclusive paths: they must be combined with this cross term.

We evaluate these Fourier integrals numerically. The two main maxima occur at $\gamma_0t=-0.4743$ and $2.426$, separated by a minimum at $\gamma_0t_v\simeq0.8950$. Dividing the two complex amplitudes at $t_v$ gives
\begin{equation*}
 \frac{\xi_{\rm sc}(t_v)}{\xi_{\rm in}(t_v)}
 \simeq-0.8565+0.4263\mathrm i,
 \qquad
 \left|\frac{\xi_{\rm sc}(t_v)}{\xi_{\rm in}(t_v)}\right|\simeq0.9567,
 \qquad
 \arg\!\frac{\xi_{\rm sc}(t_v)}{\xi_{\rm in}(t_v)}\simeq2.680~\mathrm{rad}.
\end{equation*}
The amplitudes therefore have similar magnitudes and a phase difference close to $\pi$. To quantify the cancellation, the three contributions to the conditional density, in units of $\gamma_0$, are
\begin{equation*}
 \frac{|\xi_{\rm in}(t_v)|^2}{\gamma_0P_{T,F}}\simeq0.4836,
 \qquad
 \frac{|\xi_{\rm sc}(t_v)|^2}{\gamma_0P_{T,F}}\simeq0.4426,
 \qquad
 \frac{2\operatorname{Re}[\xi_{\rm in}^*(t_v)\xi_{\rm sc}(t_v)]}{\gamma_0P_{T,F}}\simeq-0.8284.
\end{equation*}
Their sum, evaluated before rounding, gives $|\widetilde\psi_F(t_v)|^2/\gamma_0\simeq0.09784$. The scattered contribution is therefore substantial at the minimum; the detected signal is small because of destructive interference, not because the atoms stop radiating. As the Gaussian input decreases, the excitation stored in the atoms continues to radiate. This more persistent collective response dominates the later lobe.

The two maxima should not be identified with separate emissions from two REMs: the complex emission amplitudes of the contributing REMs must be summed before taking the squared modulus, so their interference also shapes the temporal profile.

Similar transients occur in nanofiber ensembles described by guided multiple scattering and local nonguided loss~\cite{Su2023}. A multi-lobed pulse therefore does not, by itself, demonstrate the importance of nonguided interatomic couplings. Here, their effect is identified by comparing~F and~I for the same geometry and input pulse, with identical local atomic properties and guided interactions.

\subsection{Dependence on the input bandwidth}\label{sec:bandwidth}
\begin{figure}[!htb]
\centering\includegraphics[width=.495\textwidth]{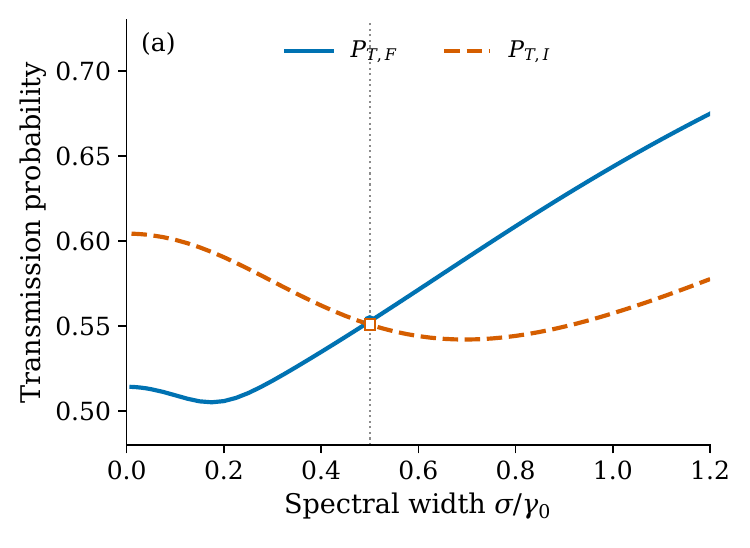}\hfill\includegraphics[width=.495\textwidth]{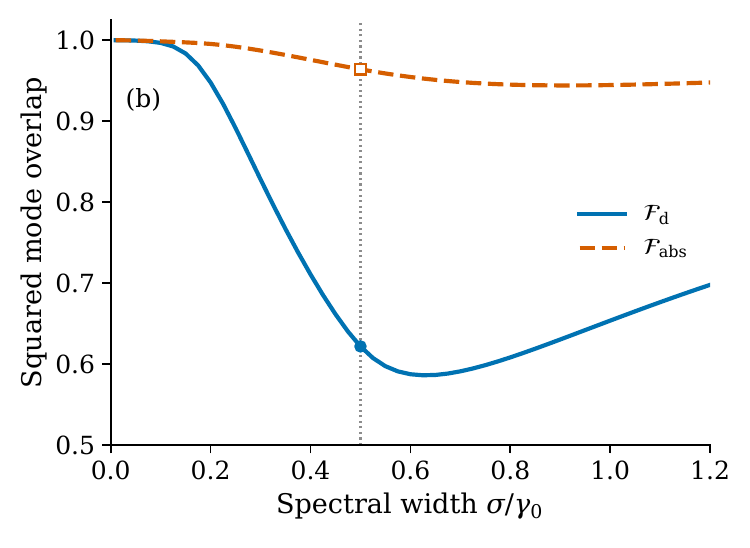}
\caption{Dependence on the input bandwidth for the chain in Eq.~\eqref{eq:parameters}, at $\delta_c=-\gamma_0$. (a) Integrated transmission probabilities. (b) Squared mode overlap after delay optimization, ${\mathcal{F}}_{\rm d}$, and overlap of the spectral moduli, ${\mathcal{F}}_{\rm abs}$. The vertical lines and symbols mark $\sigma=0.5\gamma_0$, used in Table~\ref{tab:probabilities} and Fig.~\ref{fig:modes}. The relative delay is reoptimized at each width. Blue solid and orange dashed curves in~(a) denote~F and~I. In~(b), the blue solid curve is $\mathcal F_{\rm d}$ and the orange dashed curve is $\mathcal F_{\rm abs}$. Their separation exposes phase differences that cannot be removed by a delay.}
\label{fig:bandwidth}
\end{figure}

Figure~\ref{fig:bandwidth} shows the transmission probabilities and squared mode overlaps as functions of $\sigma$ at fixed $\delta_c=-\gamma_0$. The delay is optimized for each input. The example above lies near the transmission-curve crossing, where the squared mode overlap remains well below unity.

The converse occurs at $\sigma=0.05\gamma_0$: $P_{T,F}=0.5129$ and $P_{T,I}=0.6034$, but ${\mathcal{F}}_{\rm d}=0.9996$. Across a sufficiently narrow spectrum, smooth nonzero responses differ mainly by a constant complex factor. Normalization and a global phase remove this factor, while temporal alignment removes the leading relative phase slope. As $\sigma\to0$, the modes can therefore converge while the transmission probabilities approach their generally different monochromatic limits $|t_X(\delta_c)|^2$.

\subsection{Joint detuning--bandwidth scan and scope of phase dominance}
\label{sec:scan}
\begin{figure}[!htb]
\centering
\includegraphics[width=.495\textwidth]{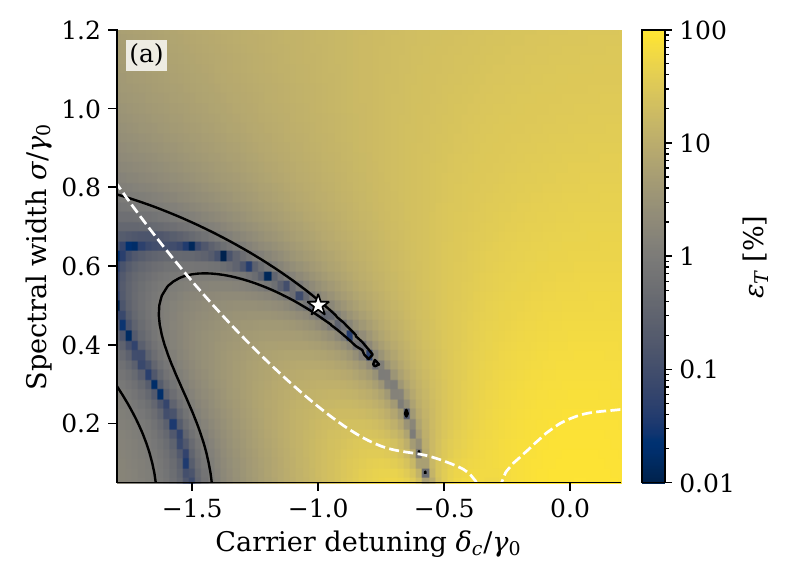}\hfill
\includegraphics[width=.495\textwidth]{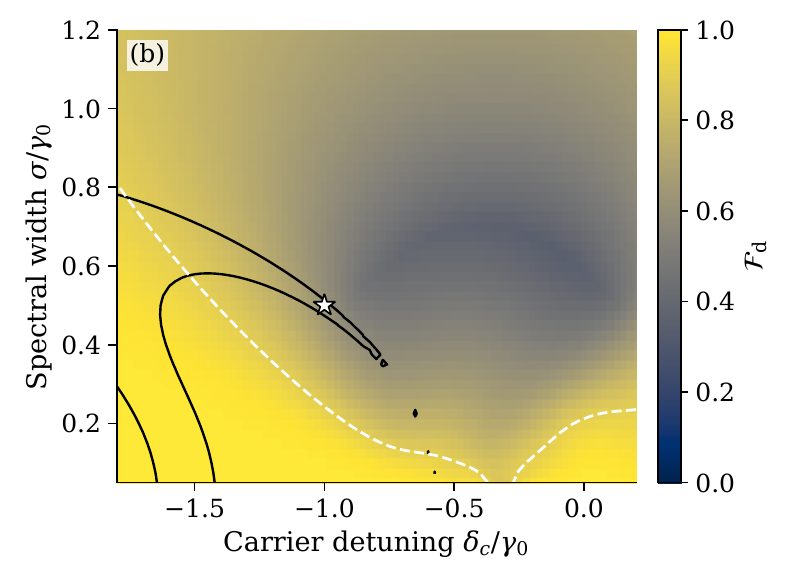}
\caption{Joint input-parameter scan at the fixed geometry of Eq.~\eqref{eq:parameters}. (a) Relative transmission discrepancy $\epsilon_T$, displayed as a percentage on a logarithmic scale. (b) Delay-optimized squared overlap $\mathcal F_{\rm d}$. The solid black contours mark $\epsilon_T=1\%$ in both panels; the white dashed contours mark $\mathcal F_{\rm d}=0.9$. The star is the original reference pulse.}
\label{fig:scan}
\end{figure}

To distinguish a finite operating region from an isolated crossing, we vary the carrier and bandwidth together, keeping the geometry fixed. Let us define the relative transmission discrepancy
\begin{equation}
 \epsilon_T=\frac{|P_{T,F}-P_{T,I}|}{P_{T,F}}.
 \label{eq:epsilon-transmission}
\end{equation}
Figure~\ref{fig:scan} displays $\epsilon_T$ and $\mathcal F_{\rm d}$ on a grid covering $-1.8\leq\delta_c/\gamma_0\leq0.2$ and $0.05\leq\sigma/\gamma_0\leq1.2$, with increments $0.025$ in both variables. The delay is optimized independently at each point. The contours $\epsilon_T=0.01$ and $\mathcal F_{\rm d}=0.9$ are plotted as illustrative criteria for close transmission and appreciable mode mismatch.

At $\sigma=0.5\gamma_0$, the condition $\epsilon_T\leq1\%$ holds approximately for $-1.065\lesssim\delta_c/\gamma_0\lesssim-0.9748$; within this interval $\mathcal F_{\rm d}$ ranges from about $0.5961$ to $0.6835$, while $\mathcal F_{\rm abs}$ remains between $0.9608$ and $0.9716$. Thus close transmission probabilities and a substantial mode mismatch coexist over a finite detuning interval, not only at the point where the transmission probabilities are exactly equal. This interval is nevertheless limited, and its width depends on the chosen transmission tolerance.

We also recomputed the Green tensor for variations of $N$, $d$, or $\rho$ at fixed input $\delta_c=-\gamma_0$, $\sigma=0.5\gamma_0$; representative results are given in Table~\ref{tab:geometry}. Both the strength of the mismatch and the balance between modulus and phase depend on geometry. For example, at $d=0.5\lambda_a$, $1-\mathcal F_{\rm abs}$ exceeds $\mathcal F_{\rm abs}-\mathcal F_{\rm d}$, whereas the opposite holds at the reference spacing. Phase dominance is therefore a property of a parameter regime, not a universal consequence of omitting radiation couplings. The joint scan also contains a point satisfying $\epsilon_T\leq1\%$ and $\mathcal F_{\rm d}\leq0.9$ for which modulus differences are at least as important as residual phase.

\begin{table}[!htb]
\centering
\caption{Geometry variations at fixed $k_aa=1.4$, $n_f=1.45$, $\delta_c=-\gamma_0$ and $\sigma=0.5\gamma_0$. Each row uses newly evaluated electromagnetic couplings for that geometry. All overlaps compare~F and~I after normalizing the transmitted modes; $\epsilon_T$ is given as a percentage. The reference geometry is the middle row with $N=10$, $d/\lambda_a=0.3$, $\rho/a=2$.}
\label{tab:geometry}
\begin{tabular}{rrrrrrr}
\hline
$N$ & $d/\lambda_a$ & $\rho/a$ & $P_{T,F}$ & $\epsilon_T(\%)$ & $\mathcal F_{\rm d}$ & $\mathcal F_{\rm abs}$\\
\hline
4 & 0.3 & 2 & 0.6978 & 9.544 & 0.9492 & 0.9796\\
8 & 0.3 & 2 & 0.5446 & 11.93 & 0.7150 & 0.9559\\
10 & 0.3 & 1.5 & 0.5199 & 42.88 & 0.7583 & 0.9156\\
10 & 0.3 & 2 & 0.5527 & 0.3400 & 0.6218 & 0.9641\\
10 & 0.3 & 2.5 & 0.6105 & 18.88 & 0.7986 & 0.9418\\
10 & 0.4 & 2 & 0.6045 & 12.58 & 0.6220 & 0.9818\\
10 & 0.5 & 2 & 0.3539 & 73.12 & 0.9182 & 0.9527\\
10 & 0.6 & 2 & 0.5056 & 12.76 & 0.9915 & 0.9945\\
\hline
\end{tabular}
\end{table}

\subsection{Possible measurements and atomic implementations}
\label{sec:experiment}
We now consider the specific case of the $^{133}$Cs D$_2$ transition near $852$~nm, with $\gamma_0/2\pi\simeq5.2$~MHz, as in the nanofiber transient experiments of Ref.~\cite{Su2023}. With this choice, Eq.~\eqref{eq:parameters} gives $a\simeq190$~nm, $\rho-a\simeq190$~nm and $d\simeq256$~nm. The reference input requires $\delta_c/2\pi\simeq-5.2$~MHz and $\sigma/2\pi\simeq2.6$~MHz, corresponding to an intensity root-mean-square duration of about $31$~ns (full width at half maximum about $72$~ns). Acousto- and electro-optic modulation with time-tagged photon counting has already resolved shorter Gaussian-pulse transients on this platform~\cite{Su2023}.

The temporal intensity could be measured by repeatedly sending the same input pulse and recording each transmitted photon's arrival time relative to the pulse trigger. Dividing the counts in each time bin by the total transmitted counts and the bin width gives the conditional density $|\widetilde\psi_F(t)|^2$, after accounting for the detector response. This histogram would display the two lobes, but it contains no information about the temporal phase and therefore does not determine $\mathcal F_{\rm d}$.

To reconstruct the complex response, one possible procedure is to scan a weak, narrowband laser through the relevant frequencies and interfere its transmitted field with a phase-stable reference taken from the same laser before the nanofiber. Scanning the reference phase measures the fringe amplitude and phase at each frequency. After calibration against the empty fiber, these measurements give the modulus and phase of the sample's transmission amplitude $t(\delta)$, which can be compared with $t_F$ and $t_I$. For a known input $f(\delta)$, normalizing the measured product $f(\delta)t(\delta)$ as in Eq.~\eqref{eq:mode} gives the corresponding transmitted mode within the linear description and its overlap with the mode predicted by~I. Alternatively, actual transmitted photons could be compared with independently prepared reference photons through the two-photon interference measurement of Sec.~\ref{sec:operational}.

This numerical example does not address all aspects of an experimental realization with Cs. Deterministic occupation at $256$~nm spacing, position fluctuations and trap-induced shifts require additional control. The dipole in our model is real and radial, corresponding to a $\pi$ transition ($\Delta m_F=0$) with the quantization axis along the common radial direction. For the Cs $F=4\leftrightarrow F'=5$ hyperfine transition, excitation of one such Zeeman pair is not a closed two-level cycle: spontaneous emission can populate other ground-state magnetic sublevels. By contrast, the closed stretched-state cycle $|F=4,m_F=4\rangle\leftrightarrow|F'=5,m_F'=5\rangle$ uses a $\sigma^+$ transition with a rotating, complex dipole, rather than the real radial dipole assumed here~\cite{SteckCs}. A quantitative Cs implementation must therefore account for its multilevel structure or justify a suitable effective two-level description.

A simpler internal-level structure is offered by $^{88}$Sr on the $^1S_0\leftrightarrow{}^3P_1$ line at $689$~nm. This isotope has zero nuclear spin; its $J=0$ ground level has only $m=0$, and hence no additional ground-state Zeeman sublevels into which the optical transition can decay. A uniform bias field along the chain's common radial direction splits the three $J'=1$ excited sublevels. Addressing $m'=0$ selects a real radial $\pi$ dipole; the $m'=\pm1$ transitions must be sufficiently detuned compared with the pulse bandwidth and the relevant collective couplings and linewidths. Under that condition, the response is closer to the two-level model. The natural linewidth $\gamma_0/2\pi\simeq7.6$~kHz gives $\sigma/2\pi\simeq3.8$~kHz and an intensity root-mean-square duration of $21~\mu$s. State-insensitive trapping of Sr near a nanofiber has been demonstrated for the $|m'|=1$ components, not this $m'=0$ configuration~\cite{Kestler2023}. An implementation of the present proposal would also require an adapted trap, an ordered chain and adequate guided transmission at the probe wavelength.

\FloatBarrier
\Needspace{7\baselineskip}
\section{Collective origin of the discrepancy}
\label{sec:collective}
We now look for the collective atomic resonances responsible for the transmitted-mode discrepancy.

\subsection{Collective resonances and transmission}
The REMs are collective atomic states,
\begin{equation}
 |\Phi_{j,X}\rangle=\sum_{\ell=1}^N v_{\ell j,X}|e_\ell\rangle,
 \qquad \sum_{\ell=1}^N|v_{\ell j,X}|^2=1,
 \label{eq:mode-components}
\end{equation}
whose amplitude columns $\mathbf v_{j,X}=(v_{1j,X},\ldots,v_{Nj,X})^{\mathsf T}$ are right eigenvectors of the non-Hermitian effective Hamiltonian $\bm H_{{\rm eff},X}$~\cite{Jimenez2025}:
\begin{equation}
 \frac{\bm H_{{\rm eff},X}}{\hbar}\mathbf v_{j,X}=\lambda_{j,X}\mathbf v_{j,X},
 \qquad \lambda_{j,X}=\Omega_{j,X}-\mathrm{i}\Gamma_{j,X}/2.
 \label{eq:eigenvalues}
\end{equation}
The indices $\ell,j=1,\ldots,N$ label atoms and REMs, respectively. Here $\Omega_{j,X}$ is the collective frequency shift relative to $\omega_a$, and $\Gamma_{j,X}$ is the population-decay rate. Without an incident field, the atomic amplitudes of REM $j$ evolve with the common factor $e^{-\mathrm{i}\Omega_{j,X}t-\Gamma_{j,X}t/2}$.

Multiplying Eq.~\eqref{eq:eigenvalues} by $\mathbf v_{j,X}^\dagger$ and separating real and imaginary parts gives
\begin{equation}
 \begin{aligned}
 \Omega_{j,X}&=\mathbf v_{j,X}^\dagger\bm\Omega_X\mathbf v_{j,X},\\
 \Gamma_{j,X}&=\mathbf v_{j,X}^\dagger\bm\Gamma_X\mathbf v_{j,X}\geq0.
 \end{aligned}
 \label{eq:collective-rates}
\end{equation}
These are expectation values in the same collective state, not generally separate eigenvalues of $\bm\Omega_X$ and $\bm\Gamma_X$. The off-diagonal elements of $\bm\Omega_X$ and $\bm\Gamma_X$ contribute through products of excitation amplitudes on different atoms. The collective shifts and decay rates therefore depend on both the magnitudes and the relative phases of these amplitudes.

Using the decay matrix of model~I from Sec.~\ref{sec:models} gives
\begin{equation}
 \Gamma_{j,I}=\mathbf v_{j,I}^\dagger\bm\Gamma_g\mathbf v_{j,I}
 +\gamma_{\mathrm{rad}}\geq\gamma_{\mathrm{rad}}.
 \label{eq:floor}
\end{equation}
In model~I, every normalized collective state has radiation rate $\gamma_{\mathrm{rad}}$ and a nonnegative guided contribution. In model~F, destructive interference can reduce $\mathbf v_{j,F}^\dagger\bm\Gamma_{\mathrm{rad}}\mathbf v_{j,F}$ below $\gamma_{\mathrm{rad}}$. The total rate then falls below the bound in Eq.~\eqref{eq:floor} if the guided contribution is also sufficiently small. The coherent couplings in $\bm\Omega_{\mathrm{rad}}$ additionally modify the resonance frequencies and eigenvectors.

For the present parameters, the complex symmetric matrices $\bm H_{{\rm eff},X}$ have nondegenerate eigenvalues. Complex symmetry therefore implies~\cite{Noble2013}
\begin{equation*}
 \mathbf v_{j,X}^{\mathsf T}\mathbf v_{k,X}=q_{j,X}\delta_{jk},
 \qquad q_{j,X}=\mathbf v_{j,X}^{\mathsf T}\mathbf v_{j,X}\ne0,
\end{equation*}
where $\delta_{jk}$ is the Kronecker delta. This enables one to identify the dual rows $\widetilde{\mathbf v}_{j,X}^{\mathsf T}$, defined by $\widetilde{\mathbf v}_{j,X}^{\mathsf T}\mathbf v_{k,X}=\delta_{jk}$, as
\begin{equation}
 \widetilde{\mathbf v}_{j,X}^{\mathsf T}
 =\frac{\mathbf v_{j,X}^{\mathsf T}}{q_{j,X}}.
 \label{eq:dual-vectors}
\end{equation}
Note that since the matrices $\bm H_{{\rm eff},X}$ are generally nonnormal, their right eigenvectors need not be orthogonal under the Hermitian product, and the dual rows generally do not coincide with $\mathbf v_{j,X}^\dagger$.
The identity $\mathbb I=\sum_j\mathbf v_{j,X}\widetilde{\mathbf v}_{j,X}^{\mathsf T}$ gives
\begin{equation}
 \bm M_X^{-1}(\delta)
 =\sum_{j=1}^N\frac{\mathbf v_{j,X}\widetilde{\mathbf v}_{j,X}^{\mathsf T}}{\delta-\lambda_{j,X}}.
 \label{eq:modal-response}
\end{equation}
Substitution into Eq.~\eqref{eq:rt} gives
\begin{equation}
 \begin{aligned}
 t_X(\delta)&=1+\sum_{j=1}^N\frac{c_{j,X}}{\delta-\lambda_{j,X}},\\
 c_{j,X}&=-\frac{\mathrm{i}\gamma_g}{2}
 (\mathbf b^\dagger\mathbf v_{j,X})(\widetilde{\mathbf v}_{j,X}^{\mathsf T}\mathbf b).
 \end{aligned}
 \label{eq:rem}
\end{equation}
The coefficient $c_{j,X}$ combines excitation of REM $j$ through $\widetilde{\mathbf v}_{j,X}^{\mathsf T}\mathbf b$ and forward guided emission through $\mathbf b^\dagger\mathbf v_{j,X}$. These resonant amplitudes interfere with each other and with the directly transmitted incident amplitude, represented by 1.

\subsection{REMs accounting for the pulse distortion}
\label{sec:dominant}
We number the REMs of model~F by increasing $\Omega_{j,F}$ and compare the spectral norms of their contributions using the dimensionless weights
\begin{equation}
 W_{j,F}=\int|f(\delta)|^2\left|\frac{c_{j,F}}{\delta-\lambda_{j,F}}\right|^2{\mathrm{d}}\delta.
 \label{eq:weight}
\end{equation}
Each weight is the squared norm of one resonant contribution to the transmitted amplitude. The direct field and interference terms are omitted, so these pulse-dependent weights are not additive channel probabilities.

REMs 1, 2 and 5 have the largest weights (Table~\ref{tab:modes}). REM 4 decays most slowly but contributes weakly: lifetime alone does not determine its contribution to the pulse.

\begin{table}[!htb]
\centering
\caption{Atomic radiative eigenmodes of model~F with the three largest weights $W_{j,F}$ for the reference pulse. REM 4, which has the smallest decay rate, is included for comparison.}
\label{tab:modes}
\begin{tabular}{rrrr}
\hline
REM $j$ & $\Omega_{j,F}/\gamma_0$ & $\Gamma_{j,F}/\gamma_0$ & $W_{j,F}$\\
\hline
1 & $-0.5353$ & 1.055 & 0.2706\\
2 & $-0.5270$ & 0.3163 & 0.1208\\
5 & $-0.2448$ & 1.589 & 0.03268\\
4 & $-0.3106$ & 0.008772 & 0.001083\\
\hline
\end{tabular}
\end{table}

\begin{figure}[!htb]
\centering\includegraphics[width=.495\textwidth]{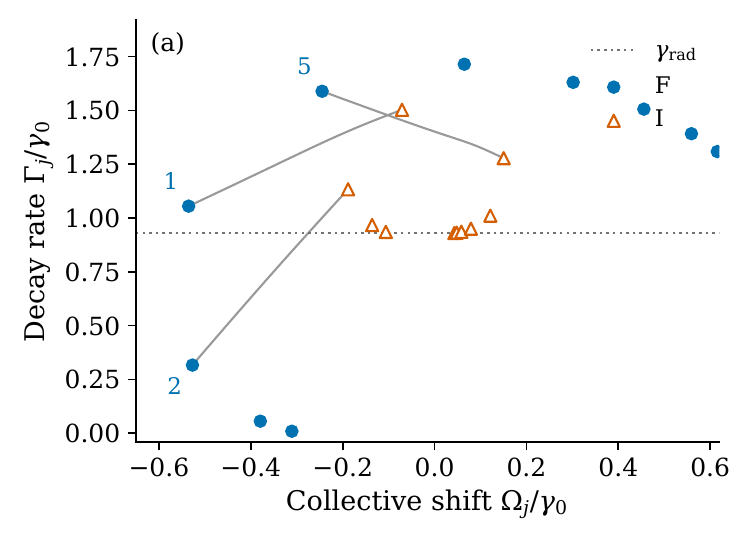}\hfill\includegraphics[width=.495\textwidth]{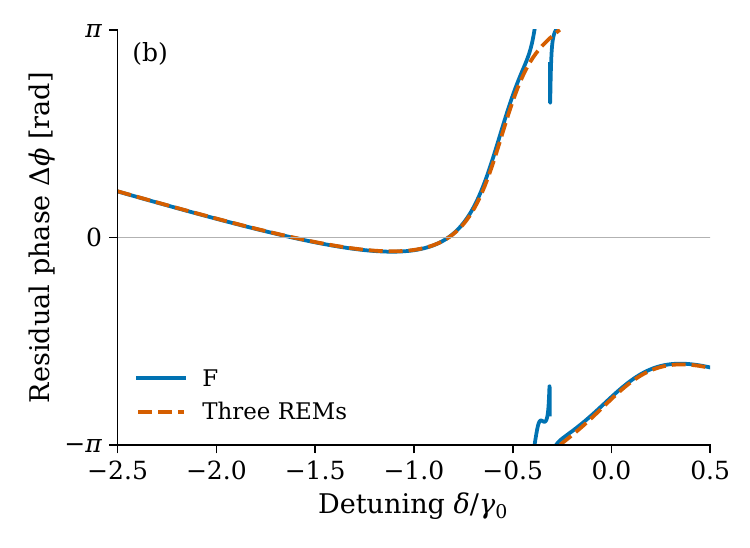}
\caption{Collective explanation for the reference chain and pulse. (a) Resonance shifts and decay rates for model~F (circles) and model~I (triangles). Labels 1, 2 and 5 identify the largest-weight REMs of model~F; curves follow them as $\eta$ increases from 0 to 1 in Eq.~\eqref{eq:interpolation}. The dotted line marks the lower bound $\gamma_{\mathrm{rad}}$ imposed by model~I. (b) Residual spectral phase relative to the response of model~I, defined in Eq.~\eqref{eq:phase}, for model~F and its reconstruction from three REMs. Both curves use the same relative delay $\tau_*$ and global phase $\phi_*$. Blue circles and orange triangles distinguish~F and~I in~(a), and thin gray curves show the interpolation rather than an extra spectrum. In~(b), the blue solid and orange dashed curves denote the full response and the three-REM reconstruction. The phase is displayed in radians in $\mathopen{]}-\pi,\pi]$; curve segments are not connected across the $2\pi$ jumps caused by wrapping at the interval boundary. These discontinuities of representation are not discontinuities of the optical field.}
\label{fig:collective}
\end{figure}

To find the counterparts of these resonances in model~I, we follow their eigenvectors continuously along the path
\begin{equation}
 \bm H_{\rm eff}(\eta)=\bm H_{{\rm eff},I}+\eta(\bm H_{{\rm eff},F}-\bm H_{{\rm eff},I}),
 \qquad 0\leq\eta\leq1.
 \label{eq:interpolation}
\end{equation}
Increasing $\eta$ restores the omitted radiation-mediated couplings. Matching normalized right eigenvectors by their Hermitian overlap at adjacent steps preserves the labels even when their frequency order changes [Fig.~\ref{fig:collective}(a)].

For REMs 1 and 2, the shift--decay pairs evolve from model~I to model~F as follows:
\begin{align*}
 (\Omega_{1,X},\Gamma_{1,X})/\gamma_0 &: (-0.07076,1.503)\longrightarrow(-0.5353,1.055),\\
 (\Omega_{2,X},\Gamma_{2,X})/\gamma_0 &: (-0.1882,1.133)\longrightarrow(-0.5270,0.3163).
\end{align*}
Their frequencies become nearly equal in model~F, while their decay rates remain very different. Their radiation contributions $\mathbf v_{j,F}^\dagger\bm\Gamma_{\mathrm{rad}}\mathbf v_{j,F}$ are $0.4011\gamma_0$ and $0.1074\gamma_0$, respectively. REM 2's total rate falls below $\gamma_{\mathrm{rad}}=0.9295\gamma_0$, the lower bound imposed by model~I.

Without fitting, we form $t_F^{(3)}(\delta)=1+\sum_{j\in\{1,2,5\}}c_{j,F}/(\delta-\lambda_{j,F})$ and normalize $f(\delta)t_F^{(3)}(\delta)$ as in Eq.~\eqref{eq:mode}.

For the phase comparison, we align $\psi_I$ with $\psi_F$ using $e^{{\mathrm{i}}\delta\tau_*+{\mathrm{i}}\phi_*}$. The residual phase and optimum global phase are
\begin{equation}
 \begin{aligned}
 \Delta\phi(\delta)&=\arg\!\left[\psi_F(\delta)\psi_I^*(\delta)
 e^{-{\mathrm{i}}\delta\tau_*-{\mathrm{i}}\phi_*}\right],\\
 \phi_*&=\arg\!\int\psi_F(\delta)\psi_I^*(\delta)
 e^{-{\mathrm{i}}\delta\tau_*}{\mathrm{d}}\delta.
 \end{aligned}
 \label{eq:phase}
\end{equation}
Here $\arg$ denotes the complex argument in radians, taken in $\mathopen{]}-\pi,\pi]$; $\Delta\phi$ is defined wherever both spectra are nonzero. Replacing $\psi_F$ in the first line by the normalized three-REM reconstruction, with the same $\tau_*$ and $\phi_*$, reproduces most of the residual phase variation [Fig.~\ref{fig:collective}(b)].

The normalized reconstruction has squared overlap $0.9879$ with the pulse from model~F at zero relative delay. After optimizing its delay relative to the pulse from model~I, the squared overlap is $0.6498$, close to $0.6218$ for model~F. These three resonances thus account for most of the discrepancy.

\subsection{Separating coherent exchange and collective decay}
\label{sec:intermediate}
The interpolation above restores both radiation-mediated exchange and collective decay. To resolve their roles, we introduce two intermediate descriptions,
\begin{equation}
 \begin{aligned}
 \bm H_{{\rm eff},C}&=\bm H_{{\rm eff},I}+\hbar\bm\Omega_{\rm rad},\\
 \bm H_{{\rm eff},D}&=\bm H_{{\rm eff},I}
 -\frac{\mathrm i\hbar}{2}(\bm\Gamma_{\rm rad}-\gamma_{\rm rad}\mathbb I).
 \end{aligned}
 \label{eq:partial-models}
\end{equation}
Relative to~I, model~C adds only the off-diagonal coherent radiation couplings, whereas~D adds only the off-diagonal radiative decay terms. Local rates and shifts and the full guided contribution remain unchanged. Both have positive semidefinite total decay matrices: that of~C is the decay matrix of~I, and that of~D is the decay matrix of~F.

\begin{table}[!htb]
\centering
\caption{Controlled separation of radiation-mediated couplings for the reference chain and pulse. Overlaps compare each normalized transmitted mode with~F, with a separately optimized delay for each comparison. The label~C denotes coherent radiation exchange only, and~D denotes collective radiative decay only, as defined in Eq.~\eqref{eq:partial-models}.}
\label{tab:intermediate}
\begin{tabular}{lrrr}
\hline
Model $X$ & $P_{T,X}$ & $\mathcal F_{\rm d}(F,X)$ & $\mathcal F_{\rm abs}(F,X)$\\
\hline
F & 0.5527 & 1 & 1\\
I & 0.5508 & 0.6218 & 0.9641\\
C & 0.3371 & 0.7526 & 0.9553\\
D & 0.7349 & 0.6896 & 0.9683\\
\hline
\end{tabular}
\end{table}

Adding both off-diagonal radiation contributions to~I recovers model~F exactly, including its complex transmission amplitude and not just its transmission probability. Table~\ref{tab:intermediate} shows that adding coherent exchange alone changes $P_T$ from $0.5508$ to $0.3371$, whereas adding collective radiative decay alone changes it to $0.7349$. Adding both gives $P_{T,F}=0.5527$, which happens to be close to $P_{T,I}$ for this pulse.

Each partial restoration therefore produces a large change in transmission, despite the small net F--I difference. Moreover, neither~C nor~D reproduces the full transmitted mode: their optimized overlaps with~F are $0.7526$ and $0.6896$, respectively. Both families of radiation-mediated couplings are needed to reproduce the reference complex response within these comparisons.

It is important to underline that the two corrections add at the Hamiltonian level,
\begin{equation*}
 \bm H_{{\rm eff},F}=\bm H_{{\rm eff},C}+\bm H_{{\rm eff},D}-\bm H_{{\rm eff},I}.
\end{equation*}
Their effects on the outgoing pulse do not add in the same way. Changing either correction changes the inverse matrix $\bm M_X^{-1}(\delta)$ in Eq.~\eqref{eq:rt}, hence the frequencies, linewidths and complex weights of the collective contributions. One must then form the total transmission amplitude, take its squared modulus and integrate over the pulse spectrum. Consequently, $P_{T,F}$ is not obtained as $P_{T,C}+P_{T,D}-P_{T,I}$; likewise, the mode overlap cannot be reconstructed by adding the separate changes of overlap.

\section{Conclusion}
We have compared the full model~F and the independent-radiation approximation~I for single-photon scattering by a nanofiber-coupled atomic chain within the fixed-atom, two-level and Markov assumptions. For the reference chain and pulse, nearly equal transmission probabilities coexist with substantially different transmitted modes even after optimal delay compensation. Three collective resonances account for most of this discrepancy, which is dominated by residual spectral-phase differences. Restoring coherent radiation exchange or collective radiative decay alone does not reproduce the full complex response in this case.

The detuning--bandwidth and geometry scans show that the magnitude of the discrepancy and the relative importance of spectral phase and modulus depend on the parameters. Conversely, narrowband inputs can yield nearly identical normalized modes despite different transmission probabilities. Transmission probability and complex mode must therefore be checked separately in mode-sensitive applications.

\appendix
\section{Evaluation, convergence and reproduction of the couplings}
\label{app:methods}
\subsection{Cylindrical Green tensor and its numerical evaluation}
\label{app:green-evaluation}
\paragraph{Analytical representation.}
We start from the cylindrical expansion of the scattering Green tensor in Appendix~A of Stourm et al.~\cite{Stourm2020}, also given in Eqs.~(A3)--(A7) of Le Kien et al.~\cite{LeKien2022}. For two atoms outside the fiber, the relevant Green's function takes the form $\mathbf G=\mathbf G_0+\mathbf G^{\rm sc}$, where $\mathbf G_0$ is the free-space tensor and $\mathbf G^{\rm sc}$ the scattering correction due to the dielectric interface. Here atomic positions share the same radius $\rho$ and azimuth, and only the radial component $G_{\rho\rho}$ is needed since the transition dipoles are assumed to point along the radial direction. Let us define the dimensionless variables
\begin{equation*}
 \bar\rho=k_a\rho,\quad \bar a=k_aa,\quad
 \zeta=k_a|z_i-z_j|,\quad h=k_z/k_a,\quad
 q=\sqrt{1-h^2},\quad q_f=\sqrt{n_f^2-h^2}.
\end{equation*}
Here $k_z$ is the longitudinal Fourier variable; $k_aq$ and $k_aq_f$ are the transverse wave numbers outside and inside the fiber. With the outgoing-wave prescription specified below, one has
\begin{equation}
 \begin{aligned}
 \frac{G_{\rho\rho}^{\rm sc}}{k_a}
 &=2\int_{\mathcal C_+^\infty}\cos(h\zeta)K_\rho(h)\,\mathrm dh,\\
 K_\rho(h)&=\frac{\mathrm i}{8\pi}\sum_{m=0}^{\infty}(2-\delta_{m0})
 \left[
 r_{MM,m}\frac{m^2}{q^2\bar\rho^2}H_m^2
 +r_{NN,m}h^2(H'_m)^2
 +2r_{MN,m}\frac{mh}{q\bar\rho}H_mH'_m
 \right].
 \end{aligned}
 \label{eq:kernel}
\end{equation}
In the equation above, $H_m=H_m^{(1)}(q\bar\rho)$, and the prime differentiates the full argument. These products contain no complex conjugation. The integer $m$ labels cylindrical angular harmonics. The coefficients $r_{MM,m}$ and $r_{NN,m}$ preserve the M or N cylindrical polarization, while $r_{MN,m}=r_{NM,m}$ describes their mixing. The M and N waves are the transverse-electric (TE$_z$, $E_z=0$) and transverse-magnetic (TM$_z$, $H_z=0$) cylindrical vector waves, respectively. In a homogeneous region of bulk wave number $k$, they are defined by~\cite{LeKien2022}
\begin{equation*}
 \mathbf M=\boldsymbol\nabla\times(\mathbf e_z\Phi_m),\qquad
 \mathbf N=k^{-1}\boldsymbol\nabla\times\mathbf M,\qquad
 \Phi_m=Z_m(k_\perp\rho)e^{\mathrm i m\phi+\mathrm i k_z z},
\end{equation*}
where $k_\perp^2=k^2-k_z^2$ and $Z_m$ is a cylindrical function. These are basis waves, not the hybrid fiber eigenmodes.

To determine the reflection coefficients, we match the fields at the fiber surface for each $m,h$ and each incident M or N polarization. Near that surface, the field produced by the exterior source in the absence of the fiber is expanded in regular waves $J_m(q\bar\rho)$. The scattered exterior field instead uses $H_m^{(1)}(q\bar\rho)$ to satisfy the outgoing condition, while the interior field uses $J_m(q_f\bar\rho)$ to remain finite on the axis. Continuity of $E_\phi,E_z,H_\phi,H_z$ at $\rho=a$ gives four linear equations for the two reflected and two transmitted amplitudes. Only the reflection coefficients enter the exterior-to-exterior tensor. Their closed expressions are given in the articles cited above; our code solves the equivalent boundary system directly.

\paragraph{Guided poles.}
The common denominator of the reflection coefficients can be written
\begin{equation*}
 D_m(h)=(U_m-V_m)(n_f^2U_m-V_m)
 -\left(\frac{mh}{\bar a}\right)^2
 \left(\frac{1}{q^2}-\frac{1}{q_f^2}\right)^2,
 \quad
 U_m=\frac{J'_m(q_f\bar a)}{q_fJ_m(q_f\bar a)},\quad
 V_m=\frac{H_m^{(1)\prime}(q\bar a)}{qH_m^{(1)}(q\bar a)}.
\end{equation*}
The characteristic equation $D_m(h)=0$ has guided roots with $1<|h|<n_f$: the corresponding exterior field is radially evanescent while propagating along the fiber. In the single-mode regime used here, the only guided family is HE$_{11}$, with $|m|=1$ and one positive pole $h_g=\beta/k_a$; the opposite direction has a pole at $-h_g$. The two polarizations are degenerate in $h_g$. Note that this restriction does not concern the radiation continuum, for which all angular orders are included until numerical convergence is reached.

\paragraph{Outgoing contour and finite numerical path.}
On the real axis, $0\leq h<1$ gives $q>0$ and outgoing radial Hankel waves, whereas $h>1$ gives $q=\mathrm i\sqrt{h^2-1}$ and radial decay. The point $h=1$ is thus the radiation threshold and a square-root branch point, not a guided pole. The outgoing solution is selected by first adding an arbitrarily small positive imaginary part to the permittivities, which attenuates waves away from their sources, and then letting this added absorption tend to zero. This is the limiting-absorption prescription~\cite{ScheelBuhmann2008}; no finite absorption is kept in the calculation. With time dependence $e^{-\mathrm i\omega t}$, the positive guided pole approaches the real axis from above, so the integration path passes below it, with $q$ continued from the outgoing branch. A lower-half-plane detour also avoids evaluating the cylindrical functions at $q=0$ and $q_f=0$ ($h=1$ and $n_f$).

The original longitudinal integral includes both signs of $h$. For the radial component and geometry considered here, the kernel is even: combining the positive and negative parts replaces $e^{\mathrm ih\zeta}+e^{-\mathrm ih\zeta}$ by $2\cos(h\zeta)$. Hence we only need to integrate along the contour $\mathcal C_+^\infty$ starting at zero and extending to positive infinity. Along its detour, $h$ is complex with nonnegative real part. Note that the use of one half of the contour is a symmetry reduction, not a restriction to photons propagating in one direction.

\begin{figure}[!htb]
\centering\includegraphics[width=0.98\textwidth]{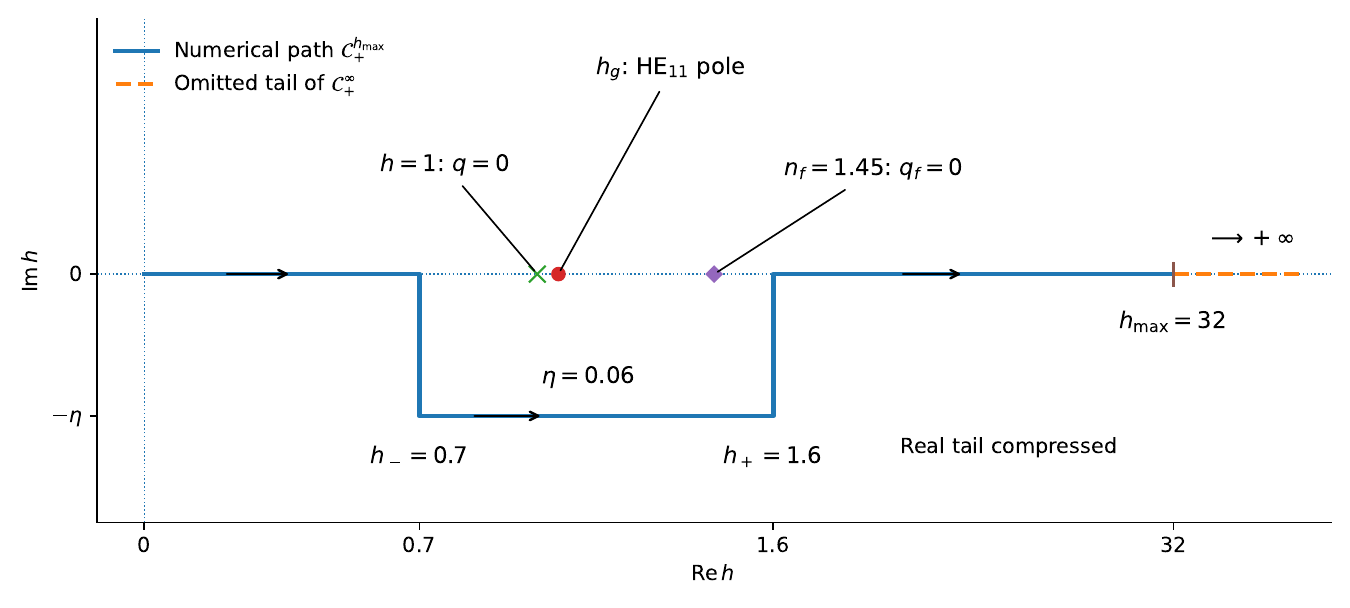}
\caption{Sommerfeld contour for the radial projection. The full path $\mathcal C_+^\infty$ consists of the solid line followed by the dashed tail to infinity. Numerical quadrature uses only the solid part, ending at $h_{\max}$. The detour passes below the exterior threshold $h=1$, the guided pole $h_g\simeq1.054$ and the point $h=n_f$ where the interior transverse wave number vanishes. The return point $h_+=1.6$ is beyond them. On a real segment at $h=x>1$, the integrand is understood as its value at $h=x-\mathrm i\varepsilon$ as $\varepsilon>0$ tends to zero. This selects $q=+\mathrm i\sqrt{x^2-1}$ and a radially decaying exterior field. The horizontal axis is compressed beyond $h=1.6$ to display the endpoint $h_{\max}=32$; this part is not drawn to scale. The pole $h_g$ is drawn at its actual position relative to 1. The negative-$h$ contribution is already included by the cosine in Eq.~\eqref{eq:kernel}.}
\label{fig:sommerfeld}
\end{figure}

Figure~\ref{fig:sommerfeld} distinguishes the infinite contour from the finite path used in the quadrature. To be more explicit, we join
\begin{equation*}
 0\ \longrightarrow\ h_-\ \longrightarrow\ h_--\mathrm i\eta
 \ \longrightarrow\ h_+-\mathrm i\eta\ \longrightarrow\ h_+
 \ \longrightarrow\ +\infty,
 \qquad h_-=0.7,\quad h_+=1.6.
\end{equation*}
The corner points delimit the straight segments of our contour. They are merely numerical choices: $h_-<1$ and $h_+>n_f=1.45$ place the detour on either side of the transverse-wave-number thresholds and the HE$_{11}$ pole. Let us underline that the depth $\eta$ is not material absorption. Deforming the path within the same analytic domain, without crossing a pole, leaves the integral unchanged.

For computation, we stop at $h_{\max}=32$, truncate the angular sum at $m_{\max}=24$, and use $\eta=0.06$. The first four integration segments join $0$, $0.7$, $0.7-\mathrm i\eta$, $1.6-\mathrm i\eta$ and $1.6$. The remaining real interval $[1.6,32]$ is divided into $[1.6,3]$, $[3,6]$, $[6,12]$ and $[12,32]$. These subdivision points are numerical choices, not additional physical thresholds. On each of the eight segments, Gauss--Legendre quadrature approximates the integral by a weighted sum of the integrand evaluated at $n_q=400$ nonuniformly spaced points, called quadrature nodes. Thus $h_{\max}$ truncates the longitudinal wave-number integral, not the interatomic distance. The effect of these numerical choices is checked in Sec.~\ref{app:accuracy}.

\paragraph{Guided contribution and residue.}
Eq.~\eqref{eq:kernel} may be computed directly via integration along the open contour in Fig.~\ref{fig:sommerfeld}. The residue at the pole may, however, be evaluated separately to identify the guided contribution already contained in this integral. In the original longitudinal representation with $e^{\mathrm ih\zeta}$, before reduction to a cosine, the guided part is isolated by deforming the contour past the positive pole: the difference between the paths before and after deformation forms a closed loop around it~\cite{Stourm2020,LeKien2022}. The residue theorem gives
\begin{equation*}
 \frac{G_{g,\rho\rho}}{k_a}=2\pi\mathrm i\mathcal R_\rho e^{\mathrm ih_g\zeta},
 \qquad \mathcal R_\rho=\lim_{h\to h_g}(h-h_g)K_\rho(h).
\end{equation*}
In the monomode regime, only $|m|=1$ contributes to this guided pole; the factor 2 in the angular sum includes both signs.

Numerically, we evaluate the kernel at $h_g\pm\epsilon$ and use the so-called Richardson extrapolation~\cite{RichardsonGaunt1927}:
\begin{equation*}
 \begin{aligned}
 \mathcal R_\rho(\epsilon)
 &=\frac{\epsilon}{2}\big[K_\rho(h_g+\epsilon)-K_\rho(h_g-\epsilon)\big]
 =\mathcal R_\rho+O(\epsilon^2),\\
 \mathcal R_\rho^{\rm ext}
 &=\frac{4\mathcal R_\rho(\epsilon/2)-\mathcal R_\rho(\epsilon)}{3}
 =\mathcal R_\rho+O(\epsilon^4).
 \end{aligned}
\end{equation*}
The combination cancels the leading quadratic error. We use $\epsilon=2\times10^{-5},10^{-5},5\times10^{-6}$ and compare the two extrapolated estimates. At coincident longitudinal coordinates, $\gamma_g/\gamma_0=6\pi\operatorname{Im}(2\pi\mathrm i\mathcal R_\rho)$; this fixes the normalization of the guided couplings in Eq.~\eqref{eq:guided}.

\paragraph{Free-space and nonguided contributions.}
For separation $R>0$, the transverse free-space contributions are evaluated analytically, with $x=k_aR$:
\begin{equation}
 \begin{aligned}
 \Omega_0(R)/\gamma_0&=-\frac34\left(\frac{\cos x}{x}-\frac{\sin x}{x^2}-\frac{\cos x}{x^3}\right),\\
 \Gamma_0(R)/\gamma_0&=\frac32\left(\frac{\sin x}{x}+\frac{\cos x}{x^2}-\frac{\sin x}{x^3}\right).
 \end{aligned}
 \label{eq:vacuum-methods}
\end{equation}
We form $\Omega_F/\gamma_0=\Omega_0/\gamma_0-3\pi\operatorname{Re}(G_{\rho\rho}^{\rm sc}/k_a)$ and $\Gamma_F/\gamma_0=\Gamma_0/\gamma_0+6\pi\operatorname{Im}(G_{\rho\rho}^{\rm sc}/k_a)$, then subtract the guided contributions to obtain $\Omega_{\rm rad}$ and $\Gamma_{\rm rad}$. At $R=0$, the common real shift is absorbed into $\omega_a$ and the free-space decay is $\gamma_0$.

\subsection{Reference values and numerical accuracy}
\label{app:accuracy}
\paragraph{Reference couplings.}
Table~\ref{tab:couplings} lists the distinct radiative-mode-assisted coupling coefficients for the chain of Eq.~\eqref{eq:parameters}. For an arbitrary atom pair $(i,j)$, use row $s=|i-j|$, corresponding to the separation $R=sd$. Together with
\begin{equation*}
 \frac{\beta}{k_a}=1.053792851780,
 \qquad \frac{\gamma_g}{\gamma_0}=0.127948403406,
 \qquad \frac{\gamma_{\rm rad}}{\gamma_0}=0.929548808096,
\end{equation*}
these entries specify $\bm H_{{\rm eff},F}$ and $\bm H_{{\rm eff},I}$ through Eqs.~\eqref{eq:guided} and~\eqref{eq:models}, then the scattering amplitudes and overlaps through Eqs.~\eqref{eq:rt}--\eqref{eq:abs}, without reevaluating the Green tensor.

Note that here $\omega_a=\omega_{\rm fs}+\Delta_{\rm fib}$, where $\omega_{\rm fs}$ includes the free-space Lamb shift, and $\Delta_{\rm fib}$ is the common single-atom transition shift induced by the fiber at radius $\rho$. The zero diagonal shift fixes the frequency origin; it does not imply that the fiber-induced shift vanishes. Using $\omega_{\rm fs}$ instead would require adding $\Delta_{\rm fib}\mathbb I$ to $\bm H_{{\rm eff},X}/\hbar$ and replacing $\delta$ by $\delta+\Delta_{\rm fib}$ in both models.

\begin{table}[!htb]
\centering
\caption{Radiative-mode-assisted coupling coefficients for the reference chain, in units of $\gamma_0$. The nonguided contribution includes free-space propagation and its modification by the fiber. The zero in row $s=0$ is set by the frequency-reference convention explained in the text.}
\label{tab:couplings}
\begin{tabular}{rrr}
\hline
$s=|i-j|$ & $\Omega_{{\rm rad},ij}/\gamma_0$ & $\Gamma_{{\rm rad},ij}/\gamma_0$\\
\hline
0 & $0.000000000000$ & $0.929548808096$\\
1 & $0.245808798993$ & $0.433603095875$\\
2 & $0.130999725595$ & $-0.227463445617$\\
3 & $-0.086028675902$ & $-0.145308737824$\\
4 & $-0.036779243541$ & $0.147236737646$\\
5 & $0.064482118762$ & $0.027095657502$\\
6 & $-0.002144060958$ & $-0.110813188825$\\
7 & $-0.045891115731$ & $0.025867532582$\\
8 & $0.020131045015$ & $0.072366837386$\\
9 & $0.026662402239$ & $-0.048900431529$\\
\hline
\end{tabular}
\end{table}

\paragraph{Convergence of the couplings.}
We test the Green-tensor calculation by increasing the angular-order limit $m_{\max}$ from 24 to 32, the number of quadrature points per segment $n_q$ from 400 to 640, and the integration limit $h_{\max}$ from 32 to 40, and by changing the contour depth $\eta$ from 0.06 to 0.10. Each parameter is varied with the others fixed, then all four are changed together. Across the ten separations in Table~\ref{tab:couplings}, the coupling coefficients change by less than $10^{-13}\gamma_0$.

\paragraph{Pulse integrals and delay optimization.}
The parameter scans in Fig.~\ref{fig:scan} use $32\,768=2^{15}$ equally spaced frequencies over $-14\leq\delta/\gamma_0\leq12$. We calculate $\mathcal F(\tau)$ on a delay grid from the squared modulus of the Fourier transform of $\psi_F^*\psi_I$. We select the largest value in $|\gamma_0\tau|\leq40$, then evaluate successive trial delays near this maximum, without restricting them to the grid. For 33 selected pulses, extending the search to $|\gamma_0\tau|\leq80$ changes $\mathcal F_{\rm d}$ by less than $6\times10^{-14}$.

Transmission probabilities and overlaps for the reference pulse and Table~\ref{tab:intermediate} are also evaluated by adaptive Gauss--Kronrod quadrature, which subdivides intervals where its estimated integration error is too large. We place subinterval boundaries at $\Omega_{j,X}$ and $\Omega_{j,X}\pm\Gamma_{j,X}/2$, the centre and half-maximum frequencies of the squared modulus of each isolated resonant factor in $\bm M_X^{-1}(\delta)$. This assigns dedicated integration intervals even to the narrowest atomic resonance, of width $0.008772\gamma_0$. The requested absolute and relative tolerances for the integration-error estimate are both $2\times10^{-11}$.

\paragraph{Numerical accuracy.}
For the reference pulse, comparing uniform grids of $65\,536$ and $131\,072$ frequencies with adaptive integration gives differences below $4\times10^{-9}$ in $\mathcal F_{\rm abs}$ and $2\times10^{-7}$ in $\gamma_0\tau_*$. Based on these comparisons and the coupling-stability checks, we estimate absolute numerical uncertainties of $10^{-7}$ for the reference probabilities and squared overlaps, and $10^{-6}/\gamma_0$ for $\tau_*$. These empirical estimates include margins above the observed changes. The transmission difference $1.879\times10^{-3}$ hence appears to be much larger than the probability uncertainty estimate.

For Fig.~\ref{fig:scan}, doubling the frequency sampling to $65\,536$ points was checked at 149 input-parameter pairs, including all 106 satisfying $\epsilon_T\leq0.01$ and $\mathcal F_{\rm d}\leq0.9$. The largest observed change in a squared overlap was $1.4\times10^{-6}$. On this basis, we use an empirical uncertainty estimate of $10^{-5}$ for the squared mode overlaps in the parameter scan, with a margin above the changes observed in these checks.

At $\sigma=0.5\gamma_0$, the interval endpoints in Sec.~\ref{sec:scan} are located by bracketing each crossing of $\epsilon_T=0.01$ and recalculating the transmission probabilities at trial detunings between the map points. The root search uses an absolute tolerance of $10^{-12}$ in $\delta_c/\gamma_0$.

\section{Derivation of the scattering amplitudes and flux balance}
\label{app:scattering}
This appendix derives Eqs.~\eqref{eq:solve}--\eqref{eq:flux} from the atom--field equations of Sec.~\ref{sec:couplings}. All operators and field amplitudes are in the frame rotating at $\omega_a$. The same derivation applies to $X=F,I$, with the corresponding effective Hamiltonian and radiation-decay matrix.

\subsection{Atomic excitation amplitudes}
\label{app:atomic-response}
The state $|G;0\rangle$ is stationary under the action of $\hat H_{\rm eff}$, which acts only on the atoms and satisfies $\hat H_{\rm eff}|G\rangle=0$. This ground--vacuum state is also stationary under the effective resonant atom--field dynamics, which conserves excitation number. With our convention $\hat\sigma_i^z=|g\rangle_i\langle g|-|e\rangle_i\langle e|$, it satisfies
\begin{equation*}
 \langle G;0|\hat\sigma_i^z(t)=\langle G;0|.
\end{equation*}
Taking the matrix element of Eq.~\eqref{eq:atomic-operators} between $\langle G;0|$ and $|\Psi_{\rm in}\rangle$ consequently gives
\begin{equation}
 \dot\alpha_{i,X}(t)
 =-\mathrm i\sum_j\left(\Omega_{ij,X}-\frac{\mathrm i}{2}\Gamma_{ij,X}\right)\alpha_{j,X}(t)
 +\frac{\mathrm ip}{\hbar}\mathbf u\cdot
 \langle G;0|\hat{\mathbf E}_{\rm in}^{(+)}(\mathbf r_i,t)|\Psi_{\rm in}\rangle.
 \label{eq:app-amplitudes}
\end{equation}
This projection is linear without replacing the operator $\hat\sigma_i^z$ by a mean population: it uses the single-excitation input and the ground-state bra. The incident-field normalization in Sec.~\ref{sec:scattering} gives
\begin{equation}
 \dot{\bm\alpha}_X(t)
 =-\frac{\mathrm i}{\hbar}\bm H_{{\rm eff},X}\bm\alpha_X(t)
 -\mathrm i\sqrt{\frac{\gamma_g}{2}}\,\xi_{\rm in}(t)\mathbf b.
 \label{eq:app-driven}
\end{equation}
Since the Fourier transform of $\xi_{\rm in}(t)$ is $f(\delta)$, Eq.~\eqref{eq:app-driven} becomes
\begin{equation}
 \left(\delta\mathbb I-\frac{\bm H_{{\rm eff},X}}{\hbar}\right)
 \bar{\bm\alpha}_X(\delta)
 =\sqrt{\frac{\gamma_g}{2}}\,f(\delta)\mathbf b.
 \label{eq:app-stationary}
\end{equation}
Inverting the matrix on the left yields Eq.~\eqref{eq:solve}.

\subsection{Projection onto the forward and backward guided modes}
\label{app:guided-amplitudes}
Choose an observation point $\mathbf r=(\rho,0,z)$ outside the chain, at the same transverse position as the atoms. The guided Green tensor then has the radial projection
\begin{equation}
 \frac{\mu_0\omega_a^2p^2}{\hbar}\,
 \mathbf u\cdot\mathbf G_g(\mathbf r,\mathbf r_j,\omega_a)\cdot\mathbf u
 =\frac{\mathrm i\gamma_g}{2}e^{\mathrm i\beta|z-z_j|}.
 \label{eq:app-guided-green}
\end{equation}
This is the propagation factor and coupling normalization used in Eq.~\eqref{eq:guided}, now evaluated at the observation point. For the selected guided polarization, introduce the scalar field amplitude
\begin{equation*}
 \mathcal E_{g,X}(z,t)
 =-\frac{p}{\hbar\sqrt{\gamma_g/2}}\,
 \mathbf u\cdot\langle G;0|\hat{\mathbf E}_g^{(+)}(\rho,0,z,t)|\Psi_{\rm in}\rangle.
\end{equation*}
Its Fourier transform $\bar{\mathcal E}_{g,X}(z,\delta)$ has incident part $f(\delta)e^{\mathrm i\beta z}$. Substitution of Eq.~\eqref{eq:app-guided-green} into the guided projection of Eq.~\eqref{eq:field-input-output}, followed by Fourier transformation, gives
\begin{equation}
 \bar{\mathcal E}_{g,X}(z,\delta)
 =f(\delta)e^{\mathrm i\beta z}
 -\mathrm i\sqrt{\frac{\gamma_g}{2}}\sum_j e^{\mathrm i\beta|z-z_j|}\bar\alpha_{j,X}(\delta).
 \label{eq:app-field}
\end{equation}
To the right of the chain, $z>z_N$, the source phase is $e^{\mathrm i\beta z}e^{-\mathrm i\beta z_j}$. To the left, $z<z_1$, the emitted wave instead carries $e^{-\mathrm i\beta z}e^{\mathrm i\beta z_j}$. Factoring out the respective observation-point phases gives the transmitted and reflected spectra referred to $z=0$:
\begin{equation}
 \begin{aligned}
 f(\delta)t_X(\delta)&=f(\delta)
 -\mathrm i\sqrt{\frac{\gamma_g}{2}}\,\mathbf b^\dagger\bar{\bm\alpha}_X(\delta),\\
 f(\delta)r_X(\delta)&=-\mathrm i\sqrt{\frac{\gamma_g}{2}}\,\mathbf b^{\mathsf T}\bar{\bm\alpha}_X(\delta).
 \end{aligned}
 \label{eq:app-output}
\end{equation}
The transmitted field includes the incident wave; the reflected field does not, since there is no incident wave from the right. Substituting Eq.~\eqref{eq:solve} and identifying the coefficients multiplying $f(\delta)$ gives Eq.~\eqref{eq:rt}.

\subsection{Emission into radiation modes}
\label{app:radiation}
The radiation-decay matrix specifies the total emission rate into the nonguided channels. To make this statement explicit, we write the corresponding jump operators as $\hat L_{\mu,X}=\sum_j\ell_{\mu j,X}\hat\sigma_j^-$. Their coefficients obey
\begin{equation*}
 (\Gamma_{{\rm rad},X})_{ij}
 =\sum_{\mu\in{\rm rad}}\ell_{\mu i,X}^*\ell_{\mu j,X},
\end{equation*}
where the sum includes any continuum integrals. For model~F these are the radiation channels; for model~I they may be represented by independent local loss channels. Acting on the atomic single-excitation component, each jump produces the amplitude $\sum_j\ell_{\mu j,X}\alpha_{j,X}(t)$. Summing its squared modulus over the channels yields the radiative photon flux
\begin{equation}
 R_{{\rm rad},X}(t)
 =\sum_{\mu\in{\rm rad}}\left|\sum_j\ell_{\mu j,X}\alpha_{j,X}(t)\right|^2
 =\bm\alpha_X^\dagger(t)\bm\Gamma_{{\rm rad},X}\bm\alpha_X(t).
 \label{eq:app-radflux}
\end{equation}
There is no incident radiation-mode field to interfere with this emission. Integrating the flux over the full pulse gives the radiation-loss probability. Parseval's identity, followed by Eq.~\eqref{eq:solve}, yields
\begin{equation*}
 \begin{aligned}
 P_{L,X}&=\int R_{{\rm rad},X}(t)\,{\mathrm d}t
 =\int\bar{\bm\alpha}_X^\dagger(\delta)\bm\Gamma_{{\rm rad},X}
       \bar{\bm\alpha}_X(\delta)\,{\mathrm d}\delta\\
 &=\int |f(\delta)|^2L_X(\delta)\,{\mathrm d}\delta,
 \end{aligned}
\end{equation*}
where the frequency-resolved loss coefficient is
\begin{equation}
 L_X(\delta)=\frac{\gamma_g}{2}\mathbf b^\dagger
 [\bm M_X^{-1}(\delta)]^\dagger\bm\Gamma_{{\rm rad},X}
 \bm M_X^{-1}(\delta)\mathbf b\geq0.
 \label{eq:app-loss}
\end{equation}
The inequality follows from the positivity of the decay matrix.

\subsection{Interference and probability conservation}
\label{app:flux}
Since $b_j=e^{\mathrm i\beta z_j}$,
\begin{equation*}
 \frac{\gamma_g}{2}(b_i b_j^*+b_i^*b_j)
 =\gamma_g\cos[\beta(z_i-z_j)]=(\Gamma_g)_{ij}.
\end{equation*}
Hence $\bm\Gamma_g=(\gamma_g/2)(\mathbf b\mathbf b^\dagger+\mathbf b^*\mathbf b^{\mathsf T})$. The two terms correspond to the two directional emission terms in Eq.~\eqref{eq:app-output}.

For real $\delta$, Hermiticity of $\bm\Omega_X$ and $\bm\Gamma_X$ gives $\bm M_X-\bm M_X^\dagger=\mathrm i\bm\Gamma_X$, with $\bm\Gamma_X=\bm\Gamma_g+\bm\Gamma_{{\rm rad},X}$. The inverse matrices satisfy
\begin{equation}
 \begin{aligned}
 \bm M_X^{-1}-[\bm M_X^{-1}]^\dagger
 &=-[\bm M_X^{-1}]^\dagger(\bm M_X-\bm M_X^\dagger)\bm M_X^{-1}\\
 &=-\mathrm i[\bm M_X^{-1}]^\dagger\bm\Gamma_X\bm M_X^{-1}.
 \end{aligned}
 \label{eq:app-resolvent}
\end{equation}
This identity does not assume that $\bm M_X$ is normal. Multiplication on the left by $\mathbf b^\dagger$ and on the right by $\mathbf b$, followed by Eq.~\eqref{eq:rt}, gives
\begin{equation}
 \begin{aligned}
 2\operatorname{Re}(t_X-1)
 &=\gamma_g\operatorname{Im}(\mathbf b^\dagger\bm M_X^{-1}\mathbf b)\\
 &=-\frac{\gamma_g}{2}\mathbf b^\dagger[\bm M_X^{-1}]^\dagger
 \bm\Gamma_X\bm M_X^{-1}\mathbf b.
 \end{aligned}
 \label{eq:app-interference}
\end{equation}
The guided-mode contribution to this quadratic form is
\begin{equation}
 \begin{aligned}
 \frac{\gamma_g}{2}\mathbf b^\dagger[\bm M_X^{-1}]^\dagger
 \bm\Gamma_g\bm M_X^{-1}\mathbf b
 &=\frac{\gamma_g^2}{4}\left(
 |\mathbf b^\dagger\bm M_X^{-1}\mathbf b|^2+
 |\mathbf b^{\mathsf T}\bm M_X^{-1}\mathbf b|^2\right)\\
 &=|t_X-1|^2+|r_X|^2.
 \end{aligned}
 \label{eq:app-guided-flux}
\end{equation}
Together with Eq.~\eqref{eq:app-loss}, this proves
\begin{equation}
 2\operatorname{Re}(t_X-1)
 =-\left(|t_X-1|^2+|r_X|^2+L_X\right).
 \label{eq:app-optical-balance}
\end{equation}
The left-hand side is the interference between the directly transmitted amplitude $1$ and the forward re-emission amplitude $t_X-1$. Importantly, the guided emission alone gives $|t_X-1|^2+|r_X|^2$, not $|t_X|^2+|r_X|^2$. Finally,
\begin{equation*}
 |t_X|^2+|r_X|^2+L_X
 =1+2\operatorname{Re}(t_X-1)+|t_X-1|^2+|r_X|^2+L_X=1,
\end{equation*}
which is Eq.~\eqref{eq:flux}.

\ack{For the purpose of open access, the authors have applied a CC-BY public copyright licence to any Author Accepted Manuscript (AAM) version arising from this submission.}

\funding{This work was supported by the French National Research Agency (ANR) through the 1DOrder project (Grant No. ANR-22-CE47-0011).}

\data{The reference coupling values and numerical methods are given in Appendix~\ref{app:methods}. The numerical data underlying the figures, parameter scans and checks, together with the analysis scripts, are available from the corresponding author upon request.}


\begin{thebibliography}{99}
\setlength{\itemsep}{0pt}
\bibitem{Ward2014} Ward J M, Maimaiti A, Le V H and Nic Chormaic S 2014 Contributed Review: Optical micro- and nanofiber pulling rig \textit{Rev. Sci. Instrum.} \textbf{85} 111501 \href{https://doi.org/10.1063/1.4901098}{\nolinkurl{doi:10.1063/1.4901098}}
\bibitem{LeKien2005} Le Kien F, Dutta Gupta S, Balykin V I and Hakuta K 2005 Spontaneous emission of a cesium atom near a nanofiber: Efficient coupling of light to guided modes \textit{Phys. Rev. A} \textbf{72} 032509 \href{https://doi.org/10.1103/PhysRevA.72.032509}{\nolinkurl{doi:10.1103/PhysRevA.72.032509}}
\bibitem{LeKien2004} Le Kien F, Balykin V I and Hakuta K 2004 Atom trap and waveguide using a two-color evanescent light field around a subwavelength-diameter optical fiber \textit{Phys. Rev. A} \textbf{70} 063403 \href{https://doi.org/10.1103/PhysRevA.70.063403}{\nolinkurl{doi:10.1103/PhysRevA.70.063403}}
\bibitem{Vetsch2010} Vetsch E, Reitz D, Sagu\'e G, Schmidt R, Dawkins S T and Rauschenbeutel A 2010 Optical interface created by laser-cooled atoms trapped in the evanescent field surrounding an optical nanofiber \textit{Phys. Rev. Lett.} \textbf{104} 203603 \href{https://doi.org/10.1103/PhysRevLett.104.203603}{\nolinkurl{doi:10.1103/PhysRevLett.104.203603}}
\bibitem{Schneeweiss2013} Schneeweiss P, Dawkins S T, Mitsch R, Reitz D, Vetsch E and Rauschenbeutel A 2013 A nanofiber-based optical conveyor belt for cold atoms \textit{Appl. Phys. B} \textbf{110} 279--283 \href{https://doi.org/10.1007/s00340-012-5268-2}{\nolinkurl{doi:10.1007/s00340-012-5268-2}}
\bibitem{Gouraud2015} Gouraud B, Maxein D, Nicolas A, Morin O and Laurat J 2015 Demonstration of a memory for tightly guided light in an optical nanofiber \textit{Phys. Rev. Lett.} \textbf{114} 180503 \href{https://doi.org/10.1103/PhysRevLett.114.180503}{\nolinkurl{doi:10.1103/PhysRevLett.114.180503}}
\bibitem{Corzo2016} Corzo N V, Gouraud B, Chandra A, Goban A, Sheremet A S, Kupriyanov D V and Laurat J 2016 Large Bragg reflection from one-dimensional chains of trapped atoms near a nanoscale waveguide \textit{Phys. Rev. Lett.} \textbf{117} 133603 \href{https://doi.org/10.1103/PhysRevLett.117.133603}{\nolinkurl{doi:10.1103/PhysRevLett.117.133603}}
\bibitem{LeKien2014} Le Kien F and Rauschenbeutel A 2014 Propagation of nanofiber-guided light through an array of atoms \textit{Phys. Rev. A} \textbf{90} 063816 \href{https://doi.org/10.1103/PhysRevA.90.063816}{\nolinkurl{doi:10.1103/PhysRevA.90.063816}}
\bibitem{Stourm2020} Stourm E, Lepers M, Robert J, Nic Chormaic S, M{\o}lmer K and Brion E 2020 Spontaneous emission and energy shifts of a Rydberg rubidium atom close to an optical nanofiber \textit{Phys. Rev. A} \textbf{101} 052508 \href{https://doi.org/10.1103/PhysRevA.101.052508}{\nolinkurl{doi:10.1103/PhysRevA.101.052508}}
\bibitem{Vylegzhanin2023} Vylegzhanin A, Brown D J, Raj A, Kornovan D F, Everett J L, Brion E, Robert J and Nic Chormaic S 2023 Excitation of $^{87}$Rb Rydberg atoms to $nS$ and $nD$ states ($n\leq68$) via an optical nanofiber \textit{Optica Quantum} \textbf{1} 6 \href{https://doi.org/10.1364/OPTICAQ.498414}{\nolinkurl{doi:10.1364/OPTICAQ.498414}}
\bibitem{AsenjoGreen2017} Asenjo-Garcia A, Hood J D, Chang D E and Kimble H J 2017 Atom-light interactions in quasi-one-dimensional nanostructures: A Green's-function perspective \textit{Phys. Rev. A} \textbf{95} 033818 \href{https://doi.org/10.1103/PhysRevA.95.033818}{\nolinkurl{doi:10.1103/PhysRevA.95.033818}}
\bibitem{Kornovan2016} Kornovan D F, Sheremet A S and Petrov M I 2016 Collective polaritonic modes in an array of two-level quantum emitters coupled to an optical nanofiber \textit{Phys. Rev. B} \textbf{94} 245416 \href{https://doi.org/10.1103/PhysRevB.94.245416}{\nolinkurl{doi:10.1103/PhysRevB.94.245416}}
\bibitem{Jimenez2025} Jim\'enez-Jaimes J, Nic Chormaic S and Brion E 2025 Controlling radiative properties of circular atomic arrays through nanofiber guided modes \textit{Phys. Rev. A} \textbf{112} 053702 \href{https://doi.org/10.1103/91kf-xl7p}{\nolinkurl{doi:10.1103/91kf-xl7p}}
\bibitem{AsenjoSelective2017} Asenjo-Garcia A, Moreno-Cardoner M, Albrecht A, Kimble H J and Chang D E 2017 Exponential improvement in photon storage fidelities using subradiance and ``selective radiance'' in atomic arrays \textit{Phys. Rev. X} \textbf{7} 031024 \href{https://doi.org/10.1103/PhysRevX.7.031024}{\nolinkurl{doi:10.1103/PhysRevX.7.031024}}
\bibitem{Liao2016} Liao Z, Nha H and Zubairy M S 2016 Dynamical theory of single-photon transport in a one-dimensional waveguide coupled to identical and nonidentical emitters \textit{Phys. Rev. A} \textbf{94} 053842 \href{https://doi.org/10.1103/PhysRevA.94.053842}{\nolinkurl{doi:10.1103/PhysRevA.94.053842}}
\bibitem{Svendsen2023} Svendsen M B M and Olmos B 2023 Modified dipole-dipole interactions in the presence of a nanophotonic waveguide \textit{Quantum} \textbf{7} 1091 \href{https://doi.org/10.22331/q-2023-08-22-1091}{\nolinkurl{doi:10.22331/q-2023-08-22-1091}}
\bibitem{Pennetta2022} Pennetta R, Blaha M, Johnson A, Lechner D, Schneeweiss P, Volz J and Rauschenbeutel A 2022 Collective radiative dynamics of an ensemble of cold atoms coupled to an optical waveguide \textit{Phys. Rev. Lett.} \textbf{128} 073601 \href{https://doi.org/10.1103/PhysRevLett.128.073601}{\nolinkurl{doi:10.1103/PhysRevLett.128.073601}}
\bibitem{Cardenas2023} Cardenas-Lopez S, Solano P, Orozco L A and Asenjo-Garcia A 2023 Optical precursors in waveguide quantum electrodynamics \textit{Phys. Rev. Research} \textbf{5} 013133 \href{https://doi.org/10.1103/PhysRevResearch.5.013133}{\nolinkurl{doi:10.1103/PhysRevResearch.5.013133}}
\bibitem{Su2023} Su D, Jiang Y, Cardenas-Lopez S, Asenjo-Garcia A, Solano P, Orozco L A and Zhao Y 2023 Dynamical beats of short pulses in waveguide QED \textit{Phys. Rev. Research} \textbf{5} L042041 \href{https://doi.org/10.1103/PhysRevResearch.5.L042041}{\nolinkurl{doi:10.1103/PhysRevResearch.5.L042041}}
\bibitem{Sokolov2026} Sokolov I M and Guerin W 2026 Light propagation in atomic stratified media: breakdown of the textbook transfer-matrix method at high density \textit{J. Opt. Soc. Am. B} \textbf{43} 1106--1112 \href{https://doi.org/10.1364/JOSAB.590758}{\nolinkurl{doi:10.1364/JOSAB.590758}}
\bibitem{ScheelBuhmann2008} Scheel S and Buhmann S Y 2008 Macroscopic quantum electrodynamics---concepts and applications \textit{Acta Phys. Slovaca} \textbf{58} 675--809 \href{https://arxiv.org/abs/0902.3586}{\nolinkurl{arXiv:0902.3586}}
\bibitem{Dung2002} Dung H T, Kn\"oll L and Welsch D-G 2002 Resonant dipole-dipole interaction in the presence of dispersing and absorbing surroundings \textit{Phys. Rev. A} \textbf{66} 063810 \href{https://doi.org/10.1103/PhysRevA.66.063810}{\nolinkurl{doi:10.1103/PhysRevA.66.063810}}
\bibitem{Miroshnychenko2013} Miroshnychenko Y and M{\o}lmer K 2013 Off-resonant transitions in the collective dynamics of multilevel atomic ensembles \textit{J. Phys. B: At. Mol. Opt. Phys.} \textbf{46} 224009 \href{https://doi.org/10.1088/0953-4075/46/22/224009}{\nolinkurl{doi:10.1088/0953-4075/46/22/224009}}
\bibitem{Derouault2014} Derouault S and Bouchene M A 2014 One-photon wave packet interacting with two separated atoms in a one-dimensional waveguide: Influence of virtual photons \textit{Phys. Rev. A} \textbf{90} 023828 \href{https://doi.org/10.1103/PhysRevA.90.023828}{\nolinkurl{doi:10.1103/PhysRevA.90.023828}}
\bibitem{Dalibard1992} Dalibard J, Castin Y and M{\o}lmer K 1992 Wave-function approach to dissipative processes in quantum optics \textit{Phys. Rev. Lett.} \textbf{68} 580--583 \href{https://doi.org/10.1103/PhysRevLett.68.580}{\nolinkurl{doi:10.1103/PhysRevLett.68.580}}
\bibitem{Legero2003} Legero T, Wilk T, Kuhn A and Rempe G 2003 Time-resolved two-photon quantum interference \textit{Appl. Phys. B} \textbf{77} 797--802 \href{https://doi.org/10.1007/s00340-003-1337-x}{\nolinkurl{doi:10.1007/s00340-003-1337-x}}
\bibitem{SteckCs} Steck D A 2025 \textit{Cesium D Line Data}, revision 2.3.4 \href{https://steck.us/alkalidata/cesiumnumbers.pdf}{\nolinkurl{steck.us/alkalidata/cesiumnumbers.pdf}}
\bibitem{Kestler2023} Kestler G, Ton K, Filin D, Cheung C, Schneeweiss P, Hoinkes T, Volz J, Safronova M S, Rauschenbeutel A and Barreiro J T 2023 State-insensitive trapping of alkaline-earth atoms in a nanofiber-based optical dipole trap \textit{PRX Quantum} \textbf{4} 040308 \href{https://doi.org/10.1103/PRXQuantum.4.040308}{\nolinkurl{doi:10.1103/PRXQuantum.4.040308}}
\bibitem{Noble2013} Noble J H, Lubasch M and Jentschura U D 2013 Generalized Householder transformations for the complex symmetric eigenvalue problem \textit{Eur. Phys. J. Plus} \textbf{128} 93 \href{https://doi.org/10.1140/epjp/i2013-13093-1}{\nolinkurl{doi:10.1140/epjp/i2013-13093-1}}
\bibitem{LeKien2022} Le Kien F, Kornovan D F, Nic Chormaic S and Busch T 2022 Repulsive Casimir--Polder potentials of low-lying excited states of a multilevel alkali-metal atom near an optical nanofiber \textit{Phys. Rev. A} \textbf{105} 042817 \href{https://doi.org/10.1103/PhysRevA.105.042817}{\nolinkurl{doi:10.1103/PhysRevA.105.042817}}
\bibitem{RichardsonGaunt1927} Richardson L F and Gaunt J A 1927 The deferred approach to the limit \textit{Phil. Trans. R. Soc. Lond. A} \textbf{226} 299--361 \href{https://doi.org/10.1098/rsta.1927.0008}{\nolinkurl{doi:10.1098/rsta.1927.0008}}
\end{thebibliography}
\end{document}